\documentclass[aps,prx,10pt,amssymb,amsmath,superscriptaddress,tightenlines,twocolumn,notitlepage,longbibliography，
] {revtex4-2}
\usepackage{graphicx}
\usepackage{amsmath}
\usepackage{amssymb}
\usepackage{bm}
\usepackage{color}
\usepackage{hyperref}
\usepackage{tabularx}
\usepackage{multirow}
\usepackage{booktabs}
\usepackage{braket}
\usepackage{epstopdf}
\usepackage{makecell}

\begin{document}

\title{Interpretation of Crystal Energy Landscapes with Kolmogorov–Arnold Networks}

\author{Gen Zu}
\affiliation{Max Planck Institute for Chemical Physics of Solids, 01187, Dresden, Germany}

\author{Ning Mao}
\affiliation{Max Planck Institute for Chemical Physics of Solids, 01187, Dresden, Germany}

\author{Claudia Felser}
\affiliation{Max Planck Institute for Chemical Physics of Solids, 01187, Dresden, Germany}

\author{Yang Zhang}
\email{yangzhang@utk.edu}
\affiliation{Department of Physics and Astronomy, University of Tennessee, Knoxville, TN 37996, USA}
\affiliation{Min H. Kao Department of Electrical Engineering and Computer Science, University of Tennessee, Knoxville, Tennessee 37996, USA}

\begin{abstract}
Characterizing crystalline energy landscapes is essential to predicting thermodynamic stability, electronic structure, and functional behavior. While machine learning (ML) enables rapid property predictions, the ``black-box'' nature of most models limits their utility for generating new scientific insights. Here, we introduce Kolmogorov–Arnold Networks (KANs) as an interpretable framework to bridge this gap.  Unlike conventional neural networks with fixed activation functions, KANs employ learnable functions that reveal underlying physical relationships. We developed the Element-Weighted KAN, a composition-only model that achieves state-of-the-art accuracy in predicting formation energy, band gap, and work function across large-scale datasets. Crucially, without any explicit physical constraints, KANs uncover interpretable chemical trends aligned with the periodic table and quantum mechanical principles through embedding analysis, correlation studies, and principal component analysis. These results demonstrate that KANs provide a powerful framework with high predictive performance and scientific interpretability, establishing a new paradigm for transparent, chemistry-based materials informatics.

\end{abstract}

\maketitle

\begin{figure*}[!htbp]
    \centering
    \includegraphics[width=0.9\textwidth]{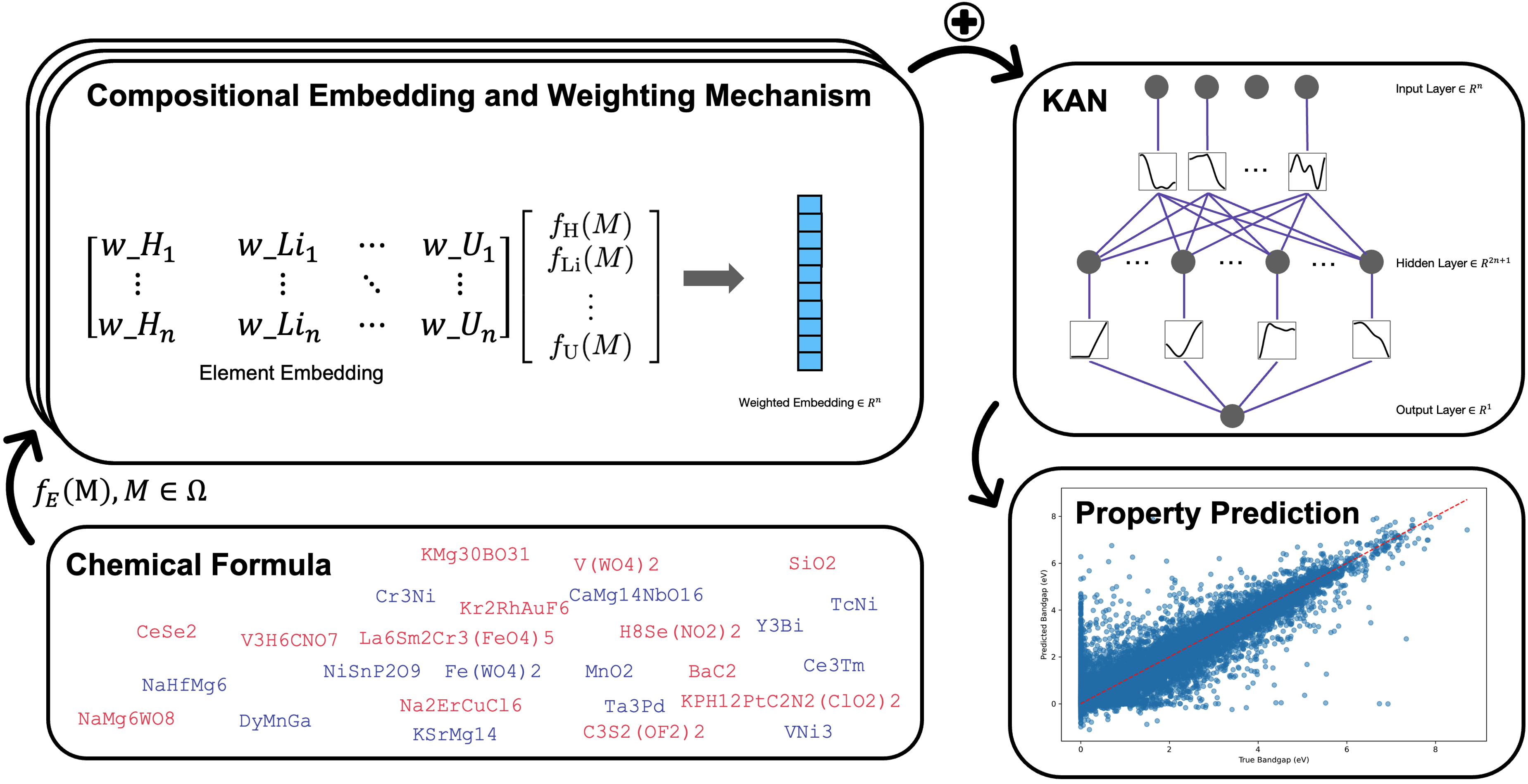}
    \caption{Workflow schematic of the Element-Weighted Kolmogorov–Arnold Network (EWKAN). Elemental compositions are transformed into weighted embeddings that capture each element’s contribution to the target property. These embeddings are passed into a two-layer KAN with trainable B-spline activation functions, which provide interpretable nonlinear mappings. The output layer yields property predictions such as band gap or formation energy.}
    \label{fig:EWKAN_schematic}
\end{figure*}

In materials science, accurately predicting the energy landscape of crystalline materials is a central challenge that underpins the discovery of new technologies, from battery materials and catalysts to semiconductors, photovoltaics and quantum devices~\cite{chen2022universal, anker2022extracting, andreeva2024potential, allan2021energy, choudhary2018machine, zuo2021accelerating}. Key properties like formation energy, band gap, dielectric constant and work function dictate a material's thermodynamic stability and functional behavior~\cite{digdaya2017interfacial, schindler2024discovery, lin2023work, weston2018machine, nassiri2021high, musil2018machine, jain2016computational, butler2018machine, wang2021vaspkit, guo2024machine}. While quantum mechanical simulations such as density functional theory (DFT) and coupled cluster (CC) provide reliable predictions, their immense computational cost prohibits the large-scale, rapid exploration of the vast chemical compound space~\cite{cohen2008insights, del2023deep, hegde2023quantifying, nakata2022large}. This limitation has motivated the development of machine learning (ML) models, which can predict material properties at a fraction of the computational expense by learning from large-scale computational and experimental databases like the Materials Project, OQMD, and AFLOW~\cite{ward2017including, schmidt2019recent, curtarolo2012aflowlib, kirklin2015open, choudhary2020joint, witman2023defect, haastrup2018computational, davariashtiyani2023formation, mao2024transfer, luo2024deep, allen2022machine, bao2025transfer}.

Despite their predictive accuracy for large-scale materials, current ML models for materials science face two critical limitations that hinder their role in scientific discovery. First, most high-performance models function as ``black-box'', relying on graph-based complex architectures that obfuscate the underlying physics and hinder researchers from extracting mechanistic insights~\cite{pilania2013accelerating, ward2016general, zhuo2018identifying, wetzel2025interpretable}. Second, most of state-of-the-art large-size models depend on graph networks of three-dimensional atomic structures as inputs, such as relaxed crystal coordinates or local atomic environments. This reliance on structural information limits their applicability in early-stage discovery, where such data may be unavailable for hypothetical compounds, or for systems like amorphous solids where they are ill-defined~\cite{du2024ctgnn, bechtel2025band, sanyal2023potential, selvaraj2024graph, merchant2023scaling}.

To overcome these challenges, our work pursues a modeling philosophy centered on two synergistic principles: compositional inputs and architectural interpretability. First, operating exclusively on chemical composition grants the model universality and speed, enabling the rapid screening of any compound defined by its chemical formula~\cite{bartel2020critical, tian2022information, antunes2022distributed,ma2023topogivity,jung2024automatic,ma2025predicting}. This approach is not only practical but also physically grounded, as composition encodes the fundamental quantum-mechanical drivers of material behavior—such as valence electron count, electronegativity, and bonding propensity~\cite{peterson2021materials, ward2016general, isayev2017universal, damewood2023representations}. 
Second, to address the ``black-box'' problem, we turn to recent advances in interpretable ML. The Kolmogorov–Arnold Network (KAN)~\cite{liu2024kan}, a new paradigm based on the Kolmogorov–Arnold representation theorem~\cite{kolmogorov1957representations}, is particularly promising in terms of expressive power and post hoc interpretability. KANs replace the fixed activation functions of traditional neural networks with learnable, univariate functions, creating a flexible and powerful architecture that allows for direct inspection of learned mathematical relationships. This architectural innovation opens promising avenues for building scientifically meaningful models in complex physical systems ~\cite{zhong2022explainable, chen2021physics, xie2023ultra, gao2025revolutionary, boura2025seqkan, ranasinghe2024ginn}.

Building on this foundation, we introduce the Element-Weighted KAN (EWKAN), a novel architecture that integrates the interpretability of KANs with a composition-only framework~\cite{huang2024material, tshitoyan2019unsupervised} for materials property prediction. Our model represents a material via its elemental constituents and their abundances, learning element-specific embeddings that are weighted and transformed through the KAN's learnable activation functions. This design allows the network to discover task-specific, nonlinear chemical representations directly from training datasets, without relying on hand-crafted features or domain-specific heuristics. We apply EWKAN to predict three critical properties in the energy landscape and functional viability of materials~\cite{sanyal2018mt, du2024ctgnn, unke2019physnet, roy2023predicting}: formation energy, band gap, and work function.

Through extensive benchmarking across large-scale datasets, the EWKAN model achieves state-of-the-art accuracy with a compact architecture. More importantly, its internal representations are chemically meaningful, aligning with periodic trends and quantum descriptors without explicit supervision. By analyzing learned activation functions and elemental embeddings through correlation and principal component analysis (PCA), we reveal how the model encodes interpretable patterns in chemical space. Our findings establish KANs as a powerful framework for transparent, composition-based materials informatics, bridging predictive performance and scientific understanding while enabling more data-efficient materials discovery.


The EWKAN architecture represents a novel approach to materials property prediction by leveraging compositional information through an adaptive activation function framework. This model integrates elemental contributions with a KAN to learn an optimal nonlinear transformation of element-weighted features.

The foundation of the model utilizes a compositional representation where each chemical element is embedded in a continuous vector space:
\begin{equation}
E = \{e_i \in \mathbb{R}^1 \mid i \in \mathcal{E}\}
\end{equation}
where $\mathcal{E}$ denotes the set of all chemical elements considered in the chosen material database, $N_e = \lvert \mathcal{E} \rvert$ denotes the total number of distinct elements, and $e_i$ represents the learned embedding for element $i$.

For a given material with composition represented as a normalized elemental distribution $\{(i, r_i)\}$, where $i$ is an element and $r_i$ is its fractional abundance, the weighted compositional representation is computed as:
\begin{equation}
f_{\text{comp}} = \sum_{i \in \mathcal{E}} r_i \cdot e_i
\end{equation}
This weighted sum aggregates the contribution of each element proportional to its abundance in the material, creating a compact representation of the material's elemental characteristics.

Rather than applying a predefined activation function such as Rectified Linear Unit (ReLU), the model employs a KAN to learn an optimal nonlinear transformation. KAN is based on the Kolmogorov-Arnold representation theorem, which is a fundamental result in function approximation theory and states that any continuous multivariate function can be represented exactly as a finite composition of continuous univariate functions and addition operations. Mathematically, the theorem establishes that for any continuous function $f: [0,1]^n \rightarrow \mathbb{R}^1$, there exists a representation:
\begin{equation}
f(\mathbf{x}) = f(x_1, \ldots, x_n) = \sum_{q=1}^{2n+1} \Phi_q\left(\sum_{p=1}^n \phi_{q,p}(x_p)\right)
\end{equation}
where $\Phi_q$ and $\phi_{q,p}$ are continuous univariate functions and $n$ denotes the dimensionality of the input space. This powerful representation theorem provides the theoretical foundation for the KAN architecture. The KAN layer implements this theorem by learning the univariate functions through parameterization. It is formulated as:
\begin{equation}
\phi(x) \;=\; w_b\, b(x) \;+\; w_s\, \mathrm{spline}(x),
\end{equation}
where \(w_b, w_s\) are trainable scalars. We choose the basis function as
\begin{equation}
b(x) \;=\; \mathrm{ReLU}(x) \;=\; \max(0, x)
\;=\; \begin{cases}
0, & x<0\\
x, & x\ge 0
\end{cases}
\end{equation}
The spline term is parameterized by B-spline basis functions:
\begin{equation}
\mathrm{spline}(x) \;=\; \sum_{l=1}^{g+k-1} c_l \, B_{l,k}(x),
\end{equation}
with trainable coefficients \(c_l\) and B-spline basis \(B_{l,k}\) of order \(k\) (degree \(k-1\)).
In our configuration, we use \(g=5\) (six grid points, five intervals) and \(k=3\)(quadratic B-splines), providing \(C^1\) continuity. The coefficients are initialized to approximate ReLU and optimized jointly with \(\{w_b, w_s, c_l\}\) during training. The KAN layer is configured with a width sequence $[n, 2n+1, 1]$, where $n$ is the dimensionality of the input feature vector, resulting in a single output dimension that represents the scalar property value. The KAN here has only two nonlinear layers, with $(2n+1)$ neurons in the hidden layer, strictly following the definition of the Kolmogorov-Arnold representation theorem. 


Therefore, the complete EWKAN model can be expressed as:
\begin{equation}
\hat{y} = \text{KAN}\left(\sum_{i \in \mathcal{E}} r_i \cdot e_i\right)
\end{equation}
where $\hat{y}$ is the predicted property value. This formulation represents the primary distinction from traditional models~\cite{ma2025predicting}, which utilize fixed activation functions such as $\text{ReLU}(\cdot)$:
\begin{equation}
\hat{y}_{\text{traditional}} = \text{ReLU}\left(\sum_{i \in \mathcal{E}} r_i \cdot e_i\right)
\end{equation}
Therefore, our approach offers several theoretical advantages. Unlike $\text{ReLU}(\cdot)$, which applies a fixed threshold-based activation, KAN learns an optimal nonlinear transformation specifically for the prediction task since its framework can approximate a wider class of functions with theoretical guarantees from the Kolmogorov-Arnold representation theorem. And in principle, the learned activation functions can be analyzed to understand the relationship between compositional features and property values to make the results more interpretable.

This architecture establishes a direct comparison between predefined activation functions and learned ones, providing insights into the importance of task-specific nonlinearities in materials property prediction.

To probe the internal representations of the KAN model, we analyze both the learned element embedding vectors and their associated activation functions. In the KAN architecture, each input passes through a learnable activation function implemented via parameterized B-spline basis functions. These activations are optimized during training and can be explicitly visualized after convergence, enabling interpretation of the nonlinear transformations applied to each embedding dimension.

To assess the physical meaning of the embedding space, we compute Pearson correlation coefficients between each embedding dimension and a curated set of elemental physicochemical properties. To uncover structural relationships among the embedding dimensions, given the embedding matrix $X \in \mathbb{R}^{N_e \times n}$, where $N_e$ is the number of elements and $n$ is the dimension of the element embedding, PCA is performed on the centered data by computing the eigendecomposition of the empirical covariance matrix:
\begin{equation}
    \Sigma = \frac{1}{n} X^\top X = V \Lambda V^\top
\end{equation}
where $V$ is the matrix of orthonormal eigenvectors (principal components) and $\Lambda$ is the diagonal matrix of eigenvalues. The projection of the embeddings into the top-$k$ components is then given by:
\begin{equation}
    Z = XV_k
\end{equation}
These projected embeddings are used for visualizing clustering behaviors and periodic trends among elements. PCA component loadings are further examined to identify which embedding dimensions contribute most to each principal direction.

To train and evaluate the proposed EWKAN model, we employ three benchmark datasets that cover essential energy-related properties of crystalline materials:
\begin{itemize}
    \item \text{Band Gap}: Band gap values are taken from the matbench\_mp\_gap dataset, a curated subset of the Materials Project database, containing 106,113 entries. These values are computed using the PBE (Perdew–Burke–Ernzerhof) exchange-correlation functional.
    \item \text{Work Function}: Work function data are obtained from a C2DB subset of the JARVIS repository, including 3,520 materials entries. The values are calculated using OPT (optimized exchange van der Waals) functional.
    \item \text{Formation Energy}: Data are sourced from the JARVIS (Joint Automated Repository for Various Integrated Simulations)-DFT database, comprising 75,993 inorganic compounds with computed formation energies (in eV/atom), also using the vdW-DF-OptB88 (OPT) functional.
\end{itemize}
Each entry in these datasets is described solely by its chemical formula, with no structural descriptors provided. This ensures that all models are trained and tested under a consistent, composition-only input regime. Preprocessing procedures, data splits, and detailed statistics for each dataset can be found in the Supplementary Materials~\ref{fig:data_bandgap} -~\ref{fig:data_formationenergy}.

The workflow of the EWKAN model is illustrated in Fig.~\ref{fig:EWKAN_schematic}. Elemental compositions are mapped to learned embeddings, which are fed into a two-layer KAN for property prediction. The dimensionality of the embedding matrix (or vector, in the case of a one-dimensional representation) is determined by the number of distinct elements in the dataset and the selected embedding dimension that reflects each element’s contribution to the target property. For each material, a weighted embedding is constructed and used as the model input. The KAN employs a trainable B-spline-based activation function optimized jointly with other model parameters.

\begin{figure}[h]
    \centering
    \includegraphics[width=\linewidth]{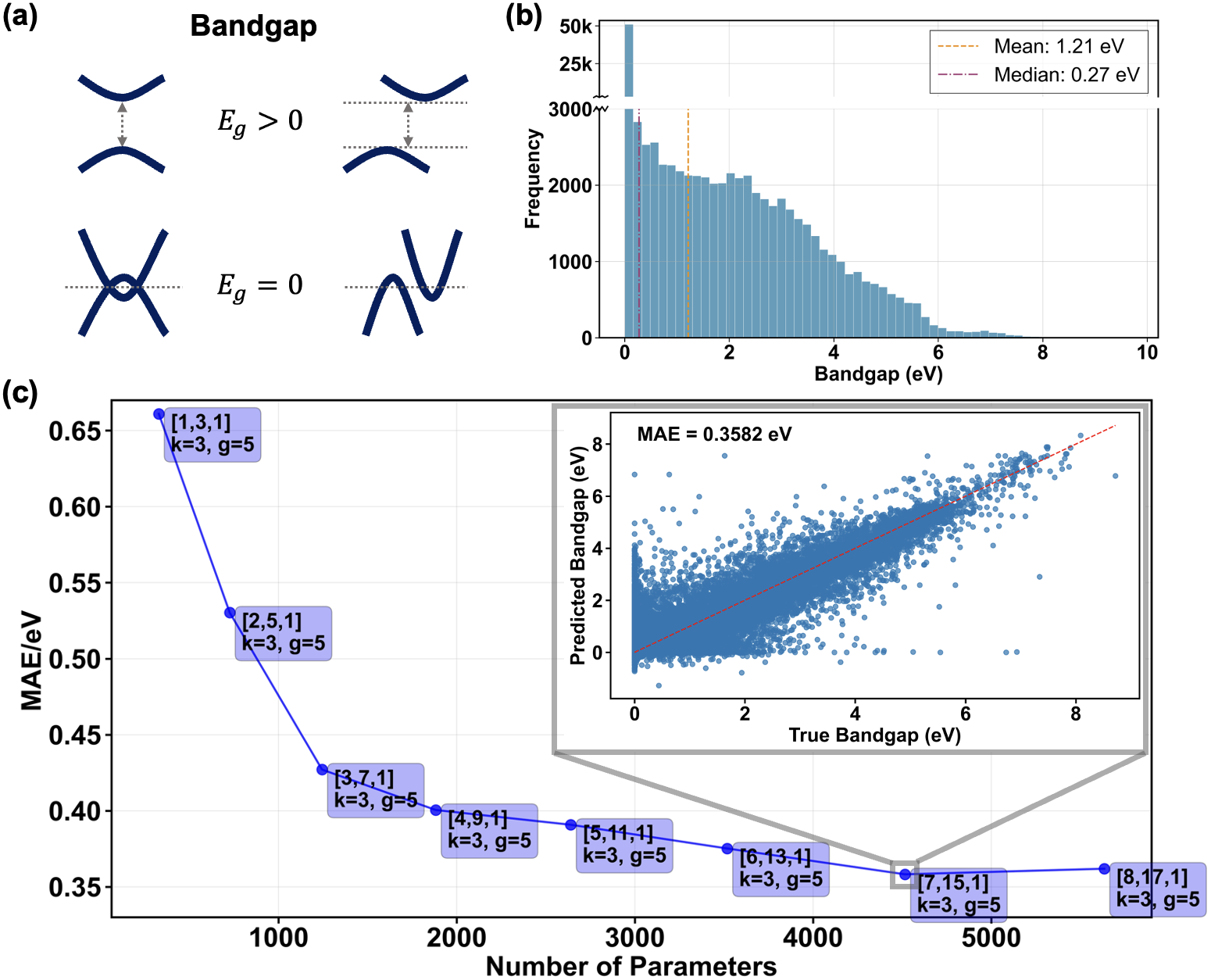}
    \caption{Bandgap prediction under EWKAN model. (a) Bandgap schematic diagram. (b) Histogram of data distribution of matbench\_mp\_gap. (c) MAE results for bandgap prediction under different KAN structure.}
    \label{fig:bandgap_accuracy}
\end{figure}

As shown in Fig.~\ref{fig:bandgap_accuracy}(a), the bandgap is defined as energy difference between the valence-band maximum and conduction-band minimum, with metals and semimetals assigned zero values. We use matbench\_mp\_gap dataset, containing 106,113 samples with bandgaps ranging from 0 to 9.721 eV; the distribution is shown in Figure.~\ref{fig:bandgap_accuracy}(b).
Based on the simple Element-Weighted model proposed by Ma et al.~\cite{ma2025predicting}, baseline results using a linear combination or $\text{ReLU}(\cdot)$ activation function are provided in the Supplementary Materials~\ref{fig:EW_matbench}. These tests yield a mean absolute error (MAE) of 0.6344 eV under 10-fold cross-validation. In addition, the Supplementary Materials~\ref{fig:periodic_table_relu} -~\ref{fig:activation_function} present visualizations of the learned elemental parameters in periodic table format~\cite{ma2023topogivity,ma2025predicting} and comparison across activation functions, which indicate that $\text{ReLU}(\cdot)$ provides the best performance within this architecture for bandgap prediction.

Although the simple Element-Weighted model uses a basic activation function, it still yields a relatively moderate MAE of 0.6344 eV. We then evaluate EWKAN with varying embedding dimensions, corresponding to different KAN architectures. As shown in Fig.~\ref{fig:bandgap_accuracy}(c), increasing embedding dimension improves predictive accuracy by effectively enlarging the model capacity. The optimal configuration, $[7,15,1]$, achieves a MAE of 0.35 eV-nearly twofold improvement over the ReLU-based baseline. Further expansion to $[8,17,1]$ yields no significant performance gain.

We evaluate model efficiency using the NetScore metric~\cite{wong2019netscore, kaplan2020scaling}, which jointly accounts for MAE and parameter count to balance accuracy and compactness. As shown in Fig.~\ref{fig:efficiency}, the $[7,15,1]$ configuration achieves the best trade-off between complexity and performance. Beyond this point, increasing the number of parameters yields diminishing returns without additional physical insight.

Analysis of the first principal component (PC1) evolution of learned activation functions provides compelling evidence for this saturation phenomenon. The variance explained by PC1 stabilizes around 54\% for configurations $[6,13,1]$ through $[8,17,1]$, while the activation function morphology converges to consistent patterns, indicating that the network has extracted all the learnable information from the underlying physics. This convergence suggests that the $[7,15,1]$ architecture captures the essential quantum mechanical principles governing the electronic band structure, with seven input dimensions corresponding to fundamental electronic descriptors for bandgap determination (detailed analysis in Supplementary Materials~\ref{fig:bandgap_pc1_evolution}). 

\begin{figure}[h]
    \centering
    \includegraphics[width=1\linewidth]{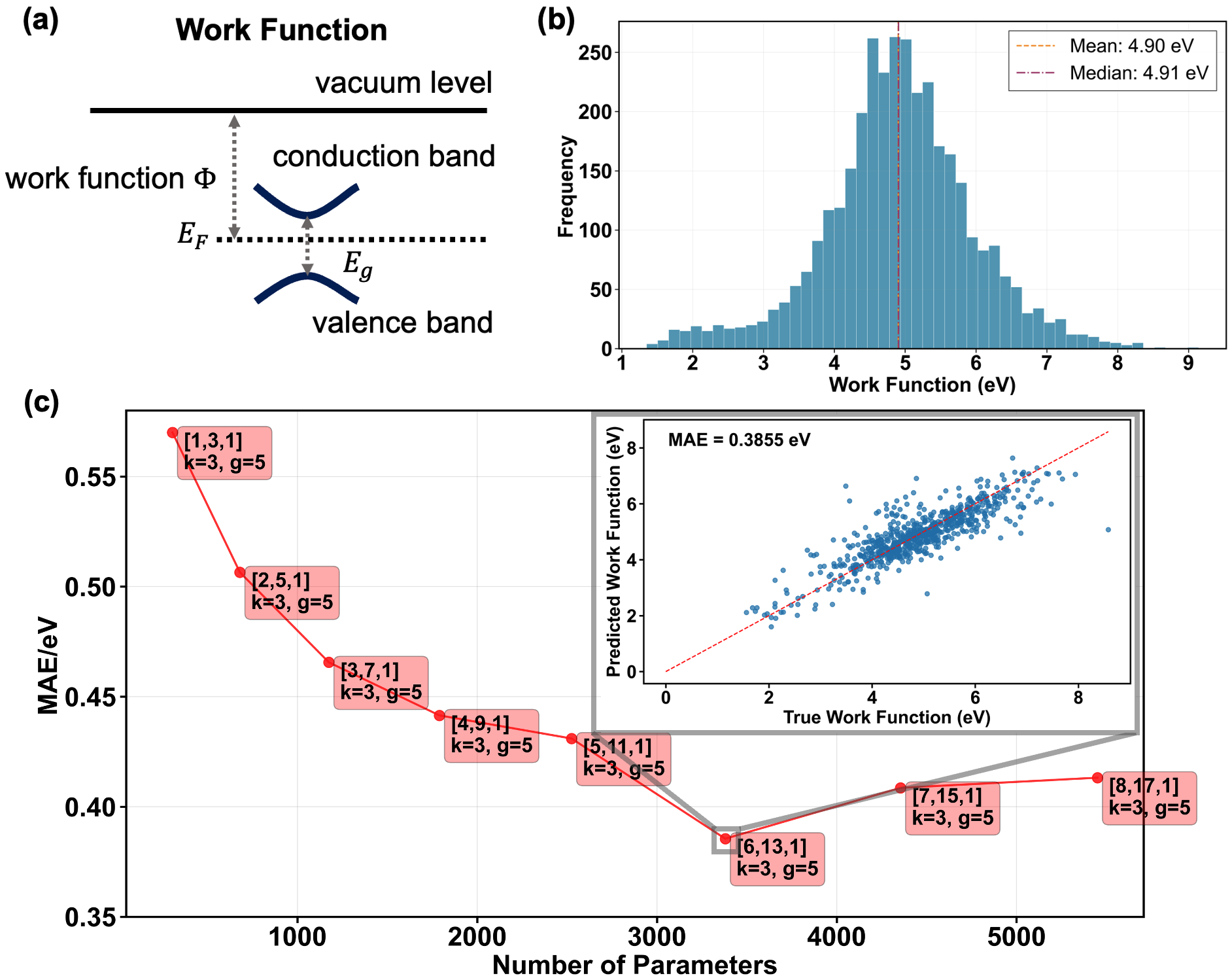}
    \caption{Work function prediction under EWKAN model. (a) Work function schematic diagram. (b) Histogram of data distribution of JARVIS-C2DB containing work function entries. (c) MAE results for work function prediction under different KAN structure.}
    \label{fig:workfunction_accuracy}
\end{figure}

As illustrated in the schematic of Fig.~\ref{fig:workfunction_accuracy}(a), the work function is defined as the minimum thermodynamic energy required to remove an electron from the Fermi level of a solid to the vacuum level just outside its surface. For this property prediction, we adopt the C2DB dataset from JARVIS for the prediction of the work function, which has 3,520 samples with values ranging from 1.348 to 9.143 eV. As shown in Fig.~\ref{fig:workfunction_accuracy}(b), the distribution of work function differs from that of the bandgap, exhibiting an approximately Gaussian profile.

As shown in Fig.~\ref{fig:workfunction_accuracy}(c), increasing the element embedding dimension and model complexity generally improves prediction accuracy. However, given the relatively limited dataset size, overly complex KAN architectures can lead to overfitting, ultimately degrading performance. The optimal performance is achieved with a KAN configuration of $[6,13,1]$, reaching a MAE of approximately 0.38 eV. Crucially, further complexity increases to $[7,15,1]$ and $[8,17,1]$ result in performance degradation rather than improvement, indicating a fundamental learning saturation. This contrasts markedly with bandgap prediction, where the optimal architecture was $[7,15,1]$, suggesting that different physical properties require distinct optimal network complexities.

PCA of the learned activation functions reveals compelling evidence for this saturation phenomenon. The PC1 explained variance 
varies within only about 1\% for configurations $[6,13,1]$ through $[8,17,1]$ (24\% to 25\%), while activation function morphologies converge to consistent patterns, indicating that the network has extracted all the learnable information about surface electronic structure governing work function. This convergence further indicates that the EWKAN model indeed captures information related to the work function (detailed analysis in Supplementary Materials~\ref{fig:workfunction_pc1_evolution}).

\begin{figure}[h]
    \centering
    \includegraphics[width=1\linewidth]{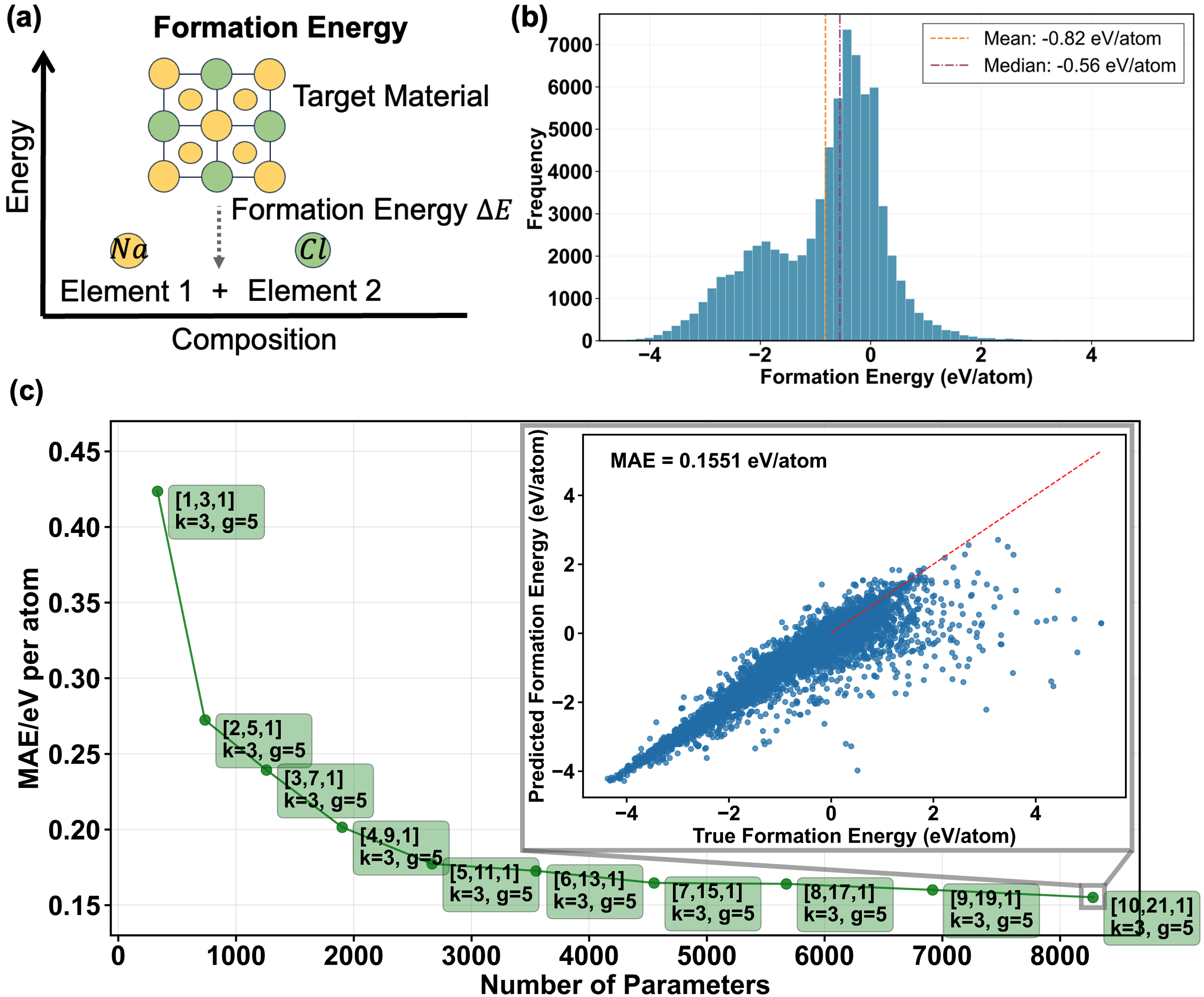}
    \caption{Formation energy prediction under EWKAN model. (a) Formation energy schematic diagram(take sodium chloride as an example). (b) Histogram of data distribution of JARVIS-DFT containing formation energy entries. (c) MAE results for formation energy prediction under different KAN structure.}
    \label{fig:formationenergy_accuracy}
\end{figure}

The JARVIS-DFT dataset is used for further prediction of formation energy, comprising 75,993 samples with values ranging from -4.42 to 5.36 eV per atom. As shown in Fig.~\ref{fig:formationenergy_accuracy}(a), the formation energy is defined as the energy difference per atom between the total energy of a compound and the sum of the chemical potentials of its constituent elements in their most stable reference states. It serves as a fundamental indicator of thermodynamic stability, where negative values imply that the formation of the compound is exothermic and energetically favorable relative to its elemental components. Fig.~\ref{fig:formationenergy_accuracy}(b) shows the empirical distribution of formation energies in the JARVIS dataset. Most materials exhibit exothermic formation energies (mean: –0.82 eV/atom; median: –0.56 eV/atom), consistent with the thermodynamic stability of crystalline phases. The relatively symmetric, unimodal distribution further facilitates statistical modeling and machine learning.

Unlike the sharp performance plateaus seen for the band gap and work function, the prediction of formation energy shows steady improvement across all tested architectures without saturation (Fig.~\ref{fig:formationenergy_accuracy}(c)). The best-performing model, with a [10,21,1] configuration, reaches a minimum MAE of 0.155 eV/atom. This continuous improvement highlights that formation energy is a more complex property than the other two. In practice, (i) formation energy reflects the combined contributions of all chemical bonds and long-range interactions; and (ii) its distribution spans a much broader range of values, requiring more model parameters to capture diverse local chemistries. In contrast, the band gap and work function depend mainly on frontier electronic states and surface dipoles, which follow lower-dimensional patterns and therefore saturate at smaller model sizes. Together, these observations explain why larger KAN architectures continue to improve formation energy predictions, indicating that this task requires greater model capacity to capture the underlying thermodynamic relationships.

This distinctive non-saturating behavior can also be interpreted through PCA. As the KAN structure increases from $[6,13,1]$ to $[8,15,1]$, the PC1 explained variance for band gap and work function changes by less than 1\%, indicating a relatively stable representation. In contrast, for formation energy, the change exceeds 6\%, reflecting a more sensitive redistribution of variance among principal components as the architecture grows. This ongoing variance fluctuation indicates that the network is possible to continue discovering new informational dimensions without reaching the learning saturation. And benefiting from a more balanced data distribution of formation energy compared to bandgap, and the larger sample size relative to work function, both the smoother MAE optimization curve and the distinctive trend in PC1 explained variance jointly demonstrate that the EWKAN model achieves its most substantial success for formation energy among the three physical quantities examined (see Supplementary Materials~\ref{fig:formationenergy_pc1_evolution}). 

To illustrate both accuracy and interpretability of our model, we examine Fe-based compounds (Table~\ref{tab:fe_compounds}). The predicted formation energies follow the expected chemical hierarchy: Fe--O compounds are the most stable (FeO: $-1.01$~eV/atom; Fe$_2$O$_3$: $-1.35$~eV/atom), Fe--S compounds are moderately stable (FeS: $-0.38$~eV/atom; FeS$_2$: $-0.39$~eV/atom), and Fe--Sn lie near zero (Fe$_3$Sn$_2$: $0.027$~eV/atom; Fe$_3$Sn: $0.075$~eV/atom). This ordering reflects established bonding principles: strong Fe--O stabilization arises from large electronegativity differences and favorable oxidation states, Fe--S shows moderate stabilization, while Fe--Sn bonding is predominantly metallic. The model thus reproduces chemically intuitive stability trends while retaining quantitative accuracy.

\begin{figure*}
    \centering
    \includegraphics[width=\textwidth]{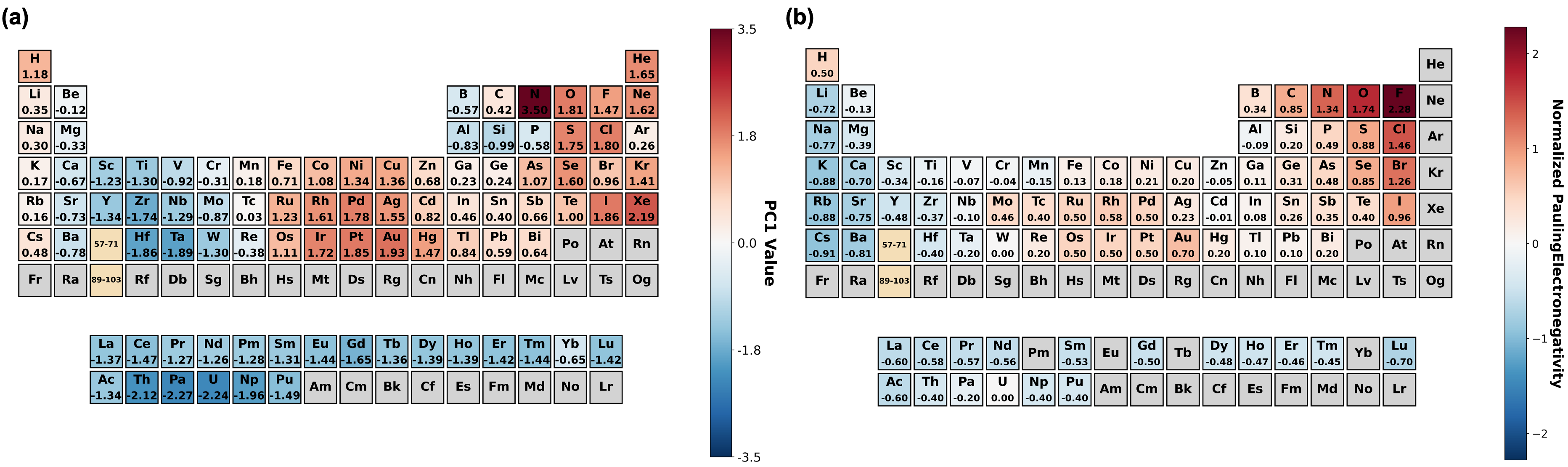}
    \caption{Comparison of (a) the original PC1 values obtained from the KAN model and (b) normalized Pauling electronegativity across the periodic table layout. The close resemblance between the two maps indicates that the dominant latent axis learned by the model captures an effective “electronegativity–metallicity” dimension, highlighting the emergence of chemically interpretable features from a purely data-driven representation. According to the PCA variance analysis (Fig.~\ref{fig:pca_var}), the PC1 achieves 32.69\% variance while the first four components account for over 80\% of the total variance}
    \label{fig:periodic_table}
\end{figure*}

Beyond predictive accuracy, the model demonstrates strong interpretability. We analyze the internal representations learned by the KAN architecture in the formation energy task. We focus on the KAN model with a $[7,15,1]$ structure, which demonstrates an effective trade-off between accuracy and complexity according to Fig.~\ref{fig:efficiency}(c). The seven-dimensional embedding sufficiently captures composition–formation energy relationships while remaining interpretable. To disentangle the distributed chemical information, we apply PCA to extract dominant variance patterns. Correlation analysis reveals that the principal components encode meaningful physicochemical properties, with strong associations to electronegativity, ionization energy, and covalent radius (Fig.~\ref{fig:formationenergy_pca}(b)).


To quantify the physical meaning behind these principal components, we compute correlations with 27 elemental properties using the Pearson correlation coefficient:
\begin{equation}
r = \frac{\sum_{i=1}^{n} (\alpha_i - \bar{\alpha})(\beta_i - \bar{\beta})}
         {\sqrt{\sum_{i=1}^{n}(\alpha_i - \bar{\alpha})^2}      \sqrt{\sum_{i=1}^{n}(\beta_i - \bar{\beta})^2}} ,
\end{equation}
where $\alpha_i$ denotes the PC1 value of the element $i$, $\beta_i$ is the corresponding elemental property value adopted from the Mendeleev database~\cite{mendeleev2014}, $\bar{\alpha}$ and $\bar{\beta}$ are their respective means. This approach, widely used in materials informatics to link latent variables with physically meaningful descriptors, provides a quantitative measure of association. The Pearson coefficient $r$ values close to $+1$ indicate a strong positive correlation, while values near $0$ imply little correlation.
Fig.~\ref{fig:formationenergy_pca}(b) reveals that PC1 strongly correlates with electronegativity ($r = 0.686$). This pattern indicates that PC1 captures the fundamental metal-nonmetal distinction that defines periodic trends.


Fig.~\ref{fig:periodic_table}(a) shows the distribution of the original PC1 values obtained from the KAN model across the periodic table, while Fig.~\ref{fig:periodic_table}(b) presents the normalized Pauling electronegativity values. The visual similarity between these two maps is striking: electropositive elements such as alkali and alkaline-earth metals exhibit strongly negative PC1 values, whereas highly electronegative elements (e.g., O, F, Cl, Br, I) are associated with large positive PC1 values. Transitioning across each period, PC1 increases in a manner that closely mirrors the canonical trend in electronegativity. This observation suggests that the primary latent axis learned by the model encodes an effective ``electronegativity–metallicity'' dimension.

The resulting organization emerges naturally from supervised learning on formation energies, without explicit knowledge of periodic structure, which means the fact that such a chemically meaningful trend emerges from a purely data-driven representation strongly supports the view that the model has learned non-trivial, physically interpretable descriptors rather than arbitrary latent features. In particular, the mapping of PC1 onto electronegativity is consistent with the underlying physics of formation energy: the driving force for compound stability is governed to a large extent by the interplay of electropositive and electronegative elements, which determines the ionic versus covalent character of bonds.

\begin{figure*}[ht]
    \centering
    \includegraphics[width=\textwidth]{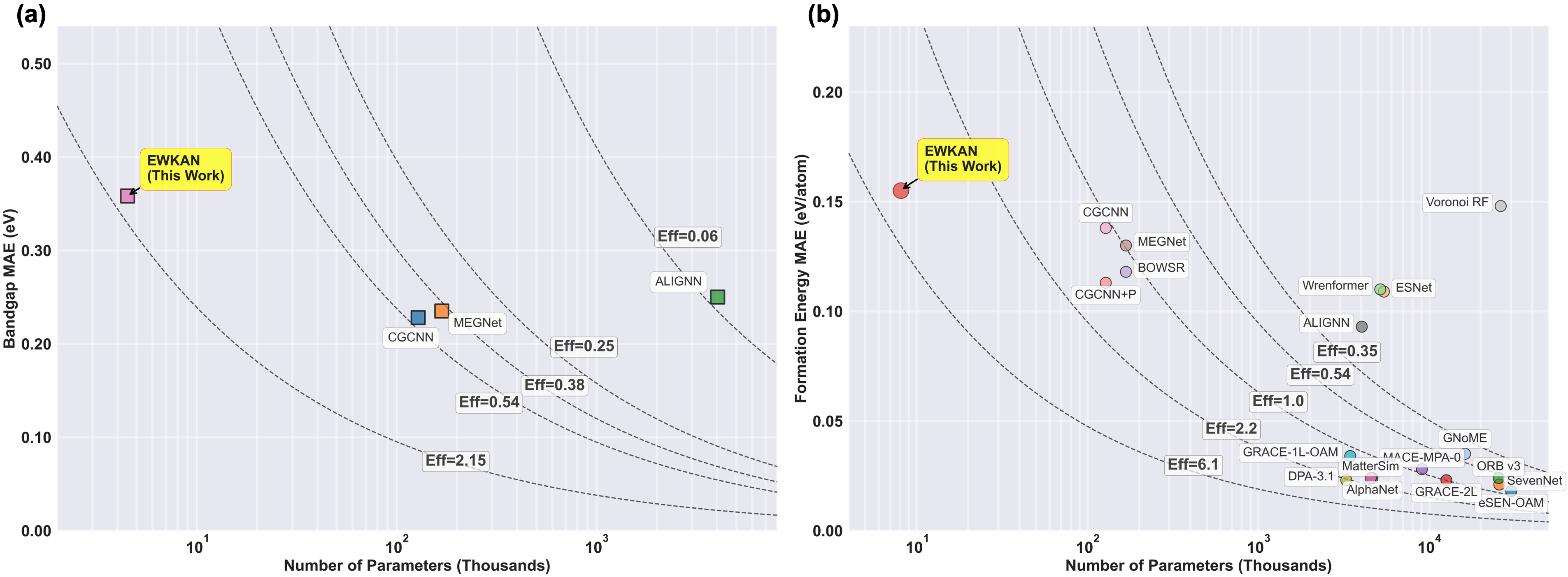}
    \caption{Comparison of model efficiency and performance across different materials property prediction tasks. (a) Mean absolute error (MAE) vs. number of parameters for band gap prediction models. (b) MAE vs. number of parameters for formation energy prediction. Dashed lines represent iso-efficiency contours, with higher curves indicating better accuracy–efficiency trade-offs. These contours are derived following the NetScore methodology \cite{wong2019netscore}, which jointly considers prediction accuracy and model complexity; see Supplementary Materials for details. EWKAN (this work) achieves competitive MAEs with significantly fewer parameters, demonstrating strong predictive efficiency in both tasks}
    \label{fig:benchmark}
\end{figure*}
In this work, we established KANs as a compelling framework for interpretable and accurate prediction of materials properties directly from chemical composition. By learning task-specific activation functions in place of fixed nonlinearities, the EWKAN model captures complex relationships between elemental makeup and physical properties such as formation energy, band gap, and work function. Across three benchmark datasets, the KAN-based approach consistently outperforms conventional models based on predefined activations, achieving competitive mean absolute errors while maintaining compact model sizes. As shown in Fig.~\ref{fig:benchmark}, our model achieves strong performance–efficiency trade-offs compared to state-of-the-art graph neural networks and large parameter models. In both band gap and formation energy prediction tasks, EWKAN occupies a favorable position in terms of model compactness and error minimization, highlighting its effectiveness for resource-efficient materials property prediction. More importantly, the learned element embeddings and activation functions encode chemically meaningful patterns—reflecting periodic trends, electronic configurations, and bonding behavior—that can be analyzed post hoc to reveal underlying physical mechanisms.

Our results show that interpretable machine learning models like KANs not only enhance predictive accuracy, but also facilitate scientific understanding by bridging data-driven learning with domain knowledge. Notably, PCA and correlation analyses of learned embeddings reveal latent representations aligned with fundamental physicochemical descriptors, suggesting that KANs can autonomously recover elemental structure–property relationships traditionally derived from quantum chemical principles.

A key limitation of our composition-only model is its inability to distinguish between chemical formula degeneracies—materials with the same chemical formula but different atomic structures. For example, the model cannot resolve the vast property differences between graphite and diamond (both pure carbon) or between metallic and molecular hydrogen, as these variations are driven by structure, not composition (see Supplementary Materials Fig.~\ref{fig:case_study}). However, this is not an intrinsic flaw of the KAN architecture but a direct consequence of the chosen input representation. This limitation highlights promising avenues for future work. The KAN framework could be readily extended to incorporate structural features—such as symmetry, coordination environment, or geometric descriptors—to resolve these degeneracies and expand its applicability (see Supplementary Materials~\ref{fig:bandgap_sg} -~\ref{fig:formation_sg}). Moreover, coupling KANs with generative models for inverse design presents a compelling route toward rational materials discovery, enabling the targeted design of stable compounds guided by interpretable, physics-informed predictions.

In conclusion, KANs provide a new paradigm for materials informatics that places understanding on equal footing with prediction. Their principled architecture, grounded in function approximation theory, makes them uniquely suited for scientific applications where transparency and understanding are as essential as accuracy. By moving from prediction to explanation, interpretable frameworks like KANs are poised to play a pivotal role in accelerating the next generation of materials design.

\section*{Supporting Information}
Details of data preparation; benchmark for ReLU; efficiency score; activation function monitoring; principal component analysis; benchmark for complicated neural network; chemical formula degeneracy; individual prediction check; Figures~\ref{fig:data_bandgap}-\ref{fig:formation_sg}; Tables~\ref{tab: EW_mae}-\ref{tab:fe_compounds}.

\section*{Acknowledge}
The authors thank Nikolai Peshcherenko and Louis Primeau for helpful discussions on model interpretability. The research by Y. Z. and N.M's visit to UT Knoxville were primarily supported by the National Science Foundation Materials Research Science and Engineering Center program through the UT Knoxville Center for Advanced Materials and Manufacturing (DMR-2309083). N.M. acknowledges the financial support from the Alexander von Humboldt Foundation. 

\section*{Data availability}
The data generated and analyzed during this study are available  from the corresponding author or at \url{https://github.com/genzuuuu/kan-energy-landscapes}.

\section*{Code availability}
All codes supporting the findings of this study are available on GitHub at \url{https://github.com/genzuuuu/kan-energy-landscapes}.

\section*{Author Contributions}
Y.Z. initiated the project. G.Z. developed the codes for machine learning and principal component analysis with assistance from N.M., and G.Z., Y.Z., and N.M. wrote the manuscript with input from all authors. All the authors participated in discussing the data and its interpretation.

\section*{Competing interests}
The authors declare no competing interests.

\section*{References}
\bibliography{references}

@article{jain2016computational,
  title={Computational predictions of energy materials using density functional theory},
  author={Jain, Anubhav and Shin, Yongwoo and Persson, Kristin A},
  journal={Nature Reviews Materials},
  volume={1},
  number={1},
  pages={1--13},
  year={2016},
  publisher={Nature Publishing Group}
}

@article{ma2023topogivity,
  title={Topogivity: A machine-learned chemical rule for discovering topological materials},
  author={Ma, Andrew and Zhang, Yang and Christensen, Thomas and Po, Hoi Chun and Jing, Li and Fu, Liang and Soljacic, Marin},
  journal={Nano Letters},
  volume={23},
  number={3},
  pages={772--778},
  year={2023},
  publisher={ACS Publications}
}

@article{butler2018machine,
  title={Machine learning for molecular and materials science},
  author={Butler, Keith T and Davies, Daniel W and Cartwright, Hugh and Isayev, Olexandr and Walsh, Aron},
  journal={Nature},
  volume={559},
  number={7715},
  pages={547--555},
  year={2018},
  publisher={Nature Publishing Group UK London}
}

@article{wang2021vaspkit,
  title={VASPKIT: A user-friendly interface facilitating high-throughput computing and analysis using VASP code},
  author={Wang, Vei and Xu, Nan and Liu, Jin-Cheng and Tang, Gang and Geng, Wen-Tong},
  journal={Computer Physics Communications},
  volume={267},
  pages={108033},
  year={2021},
  publisher={Elsevier}
}

@article{ward2017including,
  title={Including crystal structure attributes in machine learning models of formation energies via Voronoi tessellations},
  author={Ward, Logan and Liu, Ruoqian and Krishna, Amar and Hegde, Vinay I and Agrawal, Ankit and Choudhary, Alok and Wolverton, Chris},
  journal={Physical Review B},
  volume={96},
  number={2},
  pages={024104},
  year={2017},
  publisher={APS}
}

@article{pilania2013accelerating,
  title={Accelerating materials property predictions using machine learning},
  author={Pilania, Ghanshyam and Wang, Chenchen and Jiang, Xun and Rajasekaran, Sanguthevar and Ramprasad, Ramamurthy},
  journal={Scientific reports},
  volume={3},
  number={1},
  pages={2810},
  year={2013},
  publisher={Nature Publishing Group UK London}
}

@article{zhuo2018identifying,
  title={Identifying an efficient, thermally robust inorganic phosphor host via machine learning},
  author={Zhuo, Ya and Mansouri Tehrani, Aria and Oliynyk, Anton O and Duke, Anna C and Brgoch, Jakoah},
  journal={Nature communications},
  volume={9},
  number={1},
  pages={4377},
  year={2018},
  publisher={Nature Publishing Group UK London}
}

@article{curtarolo2012aflowlib,
  title={AFLOWLIB. ORG: A distributed materials properties repository from high-throughput ab initio calculations},
  author={Curtarolo, Stefano and Setyawan, Wahyu and Wang, Shidong and Xue, Junkai and Yang, Kesong and Taylor, Richard H and Nelson, Lance J and Hart, Gus LW and Sanvito, Stefano and Buongiorno-Nardelli, Marco and others},
  journal={Computational Materials Science},
  volume={58},
  pages={227--235},
  year={2012},
  publisher={Elsevier}
}

@article{choudhary2020joint,
  title={The joint automated repository for various integrated simulations (JARVIS) for data-driven materials design},
  author={Choudhary, Kamal and Garrity, Kevin F and Reid, Andrew CE and DeCost, Brian and Biacchi, Adam J and Hight Walker, Angela R and Trautt, Zachary and Hattrick-Simpers, Jason and Kusne, A Gilad and Centrone, Andrea and others},
  journal={npj computational materials},
  volume={6},
  number={1},
  pages={173},
  year={2020},
  publisher={Nature Publishing Group UK London}
}

@article{haastrup2018computational,
  title={The Computational 2D Materials Database: high-throughput modeling and discovery of atomically thin crystals},
  author={Haastrup, Sten and Strange, Mikkel and Pandey, Mohnish and Deilmann, Thorsten and Schmidt, Per S and Hinsche, Nicki F and Gjerding, Morten N and Torelli, Daniele and Larsen, Peter M and Riis-Jensen, Anders C and others},
  journal={2D Materials},
  volume={5},
  number={4},
  pages={042002},
  year={2018},
  publisher={IOP Publishing}
}

@article{kirklin2015open,
  title={The Open Quantum Materials Database (OQMD): assessing the accuracy of DFT formation energies},
  author={Kirklin, Scott and Saal, James E and Meredig, Bryce and Thompson, Alex and Doak, Jeff W and Aykol, Muratahan and R{\"u}hl, Stephan and Wolverton, Chris},
  journal={npj Computational Materials},
  volume={1},
  number={1},
  pages={1--15},
  year={2015},
  publisher={Nature Publishing Group}
}

@misc{ma2025predicting,
  author       = {{Ma, A.; Dugan, O.; Solja\v{c}i\'c, M.}},
  title        = {{Predicting band gap from chemical composition: A simple learned model for a material property with atypical statistics}},
  howpublished = {arXiv. 2025-01. \url{https://arxiv.org/abs/2501.02932}},
  note         = {(accessed 2026-03-19)}
}

@article{liu2024kan,
  title={Kan: Kolmogorov-arnold networks},
  author={Liu, Ziming and Wang, Yixuan and Vaidya, Sachin and Ruehle, Fabian and Halverson, James and Solja{\v{c}}i{\'c}, Marin and Hou, Thomas Y and Tegmark, Max},
  journal={arXiv preprint arXiv:2404.19756},
  year={2024}
}

@misc{mendeleev2014,
  author       = {Mentel, Lukasz},
  title        = {\texttt{mendeleev} -- A Python resource for properties of chemical elements, ions and isotopes},
  howpublished = {\url{https://github.com/lmmentel/mendeleev}},
  note         = {Version 1.1.0},
  year         = {2014},
}

@article{musil2018machine,
  title={Machine learning for the structure--energy--property landscapes of molecular crystals},
  author={Musil, F{\'e}lix and De, Sandip and Yang, Jack and Campbell, Joshua E and Day, Graeme M and Ceriotti, Michele},
  journal={Chemical science},
  volume={9},
  number={5},
  pages={1289--1300},
  year={2018},
  publisher={Royal Society of Chemistry}
}

@article{cohen2008insights,
  title={Insights into current limitations of density functional theory},
  author={Cohen, Aron J and Mori-S{\'a}nchez, Paula and Yang, Weitao},
  journal={Science},
  volume={321},
  number={5890},
  pages={792--794},
  year={2008},
  publisher={American Association for the Advancement of Science}
}

@article{schmidt2019recent,
  title={Recent advances and applications of machine learning in solid-state materials science},
  author={Schmidt, Jonathan and Marques, M{\'a}rio RG and Botti, Silvana and Marques, Miguel AL},
  journal={npj computational materials},
  volume={5},
  number={1},
  pages={83},
  year={2019},
  publisher={Nature Publishing Group UK London}
}

@article{zhong2022explainable,
  title={Explainable machine learning in materials science},
  author={Zhong, Xiaoting and Gallagher, Brian and Liu, Shusen and Kailkhura, Bhavya and Hiszpanski, Anna and Han, T Yong-Jin},
  journal={npj computational materials},
  volume={8},
  number={1},
  pages={204},
  year={2022},
  publisher={Nature Publishing Group UK London}
}

@article{tshitoyan2019unsupervised,
  title={Unsupervised word embeddings capture latent knowledge from materials science literature},
  author={Tshitoyan, Vahe and Dagdelen, John and Weston, Leigh and Dunn, Alexander and Rong, Ziqin and Kononova, Olga and Persson, Kristin A and Ceder, Gerbrand and Jain, Anubhav},
  journal={Nature},
  volume={571},
  number={7763},
  pages={95--98},
  year={2019},
  publisher={Nature Publishing Group}
}

@inproceedings{kolmogorov1957representations,
  title={On the representations of continuous functions of many variables by superposition of continuous functions of one variable and addition},
  author={Kolmogorov, Andrei Nikolaevich},
  booktitle={Dokl. Akad. Nauk USSR},
  volume={114},
  pages={953--956},
  year={1957}
}

@article{gao2025revolutionary,
  title={A revolutionary neural network architecture with interpretability and flexibility based on Kolmogorov--Arnold for solar radiation and temperature forecasting},
  author={Gao, Yuan and Hu, Zehuan and Chen, Wei-An and Liu, Mingzhe and Ruan, Yingjun},
  journal={Applied Energy},
  volume={378},
  pages={124844},
  year={2025},
  publisher={Elsevier}
}

@misc{boura2025seqkan,
  author       = {{Boura, T.; Konstantopoulos, S.}},
  title        = {{seqKAN: Sequence processing with Kolmogorov-Arnold Networks}},
  howpublished = {arXiv. 2025-02. \url{https://arxiv.org/abs/2502.14681}},
  note         = {(accessed 2026-03-19)}
}

@misc{ranasinghe2024ginn,
  author       = {{Ranasinghe, N.; Xia, Y.; Seneviratne, S.; Halgamuge, S.}},
  title        = {{GINN-KAN: Interpretability pipelining with applications in Physics Informed Neural Networks}},
  howpublished = {arXiv. 2024-08. \url{https://arxiv.org/abs/2408.14780}},
  note         = {(accessed 2026-03-19)}
}

@misc{andreeva2024potential,
  author       = {{Andreeva, N.A.; Chaban, V.V.}},
  title        = {{Potential Energy Landscape as a Framework for Developing Innovative Materials}},
  howpublished = {arXiv. 2024-11. \url{https://arxiv.org/abs/2411.03732}},
  note         = {(accessed 2026-03-19)}
}

@article{allan2021energy,
  title={Energy landscapes of perfect and defective solids: from structure prediction to ion conduction},
  author={Allan, Neil L and Conejeros, Sergio and Hart, Judy N and Mohn, Chris E},
  journal={Theoretical Chemistry Accounts},
  volume={140},
  number={11},
  pages={151},
  year={2021},
  publisher={Springer}
}

@article{choudhary2018machine,
  title={Machine learning with force-field-inspired descriptors for materials: Fast screening and mapping energy landscape},
  author={Choudhary, Kamal and DeCost, Brian and Tavazza, Francesca},
  journal={Physical review materials},
  volume={2},
  number={8},
  pages={083801},
  year={2018},
  publisher={APS}
}

@article{zuo2021accelerating,
  title={Accelerating materials discovery with Bayesian optimization and graph deep learning},
  author={Zuo, Yunxing and Qin, Mingde and Chen, Chi and Ye, Weike and Li, Xiangguo and Luo, Jian and Ong, Shyue Ping},
  journal={Materials Today},
  volume={51},
  pages={126--135},
  year={2021},
  publisher={Elsevier}
}

@misc{du2024ctgnn,
  author       = {{Du, Z.; Jin, L.; Shu, L.; Cen, Y.; Xu, Y.; Mei, Y.; Zhang, H.}},
  title        = {{CTGNN: Crystal Transformer Graph Neural Network for Crystal Material Property Prediction}},
  howpublished = {arXiv. 2024-05. \url{https://arxiv.org/abs/2405.11502}},
  note         = {(accessed 2026-03-19)}
}

@article{bechtel2025band,
  title={Band-gap regression with architecture-optimized message-passing neural networks},
  author={Bechtel, Tim and Speckhard, Daniel T and Godwin, Jonathan and Draxl, Claudia},
  journal={Chemistry of Materials},
  volume={37},
  number={4},
  pages={1358--1369},
  year={2025},
  publisher={ACS Publications}
}

@misc{sanyal2023potential,
  author       = {{Sanyal, S.; Sagotra, A.K.; Kumar, N.; Rathi, S.; Krishna, M.; Somayajula, N.; Palanisamy, D.; Ratnakar, R.R.; Sanyal, S.; Talukdar, P.; Waghmare, U.; Balachandran, J.}},
  title        = {{Potential energy surface prediction of Alumina polymorphs using graph neural network}},
  howpublished = {arXiv. 2023-01. \url{https://arxiv.org/abs/2301.12059}},
  note         = {(accessed 2026-03-19)}
}

@article{selvaraj2024graph,
  title={Graph Neural Networks Based Deep Learning for Predicting Structural and Electronic Properties},
  author={Selvaraj, Selva Chandrasekaran},
  journal={arXiv preprint arXiv:2411.02331},
  year={2024}
}

@article{jung2024automatic,
  title={Automatic prediction of band gaps of inorganic materials using a gradient boosted and statistical feature selection workflow},
  author={Jung, Son Gyo and Jung, Guwon and Cole, Jacqueline M},
  journal={Journal of Chemical Information and Modeling},
  volume={64},
  number={4},
  pages={1187--1200},
  year={2024},
  publisher={ACS Publications}
}

@article{chen2021physics,
  title={Physics-informed learning of governing equations from scarce data},
  author={Chen, Zhao and Liu, Yang and Sun, Hao},
  journal={Nature communications},
  volume={12},
  number={1},
  pages={6136},
  year={2021},
  publisher={Nature Publishing Group UK London}
}

@article{xie2023ultra,
  title={Ultra-fast interpretable machine-learning potentials},
  author={Xie, Stephen R and Rupp, Matthias and Hennig, Richard G},
  journal={npj Computational Materials},
  volume={9},
  number={1},
  pages={162},
  year={2023},
  publisher={Nature Publishing Group UK London}
}

@article{chen2022universal,
  title={A universal graph deep learning interatomic potential for the periodic table},
  author={Chen, Chi and Ong, Shyue Ping},
  journal={Nature Computational Science},
  volume={2},
  number={11},
  pages={718--728},
  year={2022},
  publisher={Nature Publishing Group US New York}
}

@article{anker2022extracting,
  title={Extracting structural motifs from pair distribution function data of nanostructures using explainable machine learning},
  author={Anker, Andy S and Kj{\ae}r, Emil TS and Juelsholt, Mikkel and Christiansen, Troels Lindahl and Skj{\ae}rv{\o}, Susanne Linn and J{\o}rgensen, Mads Ry Vogel and Kantor, Innokenty and S{\o}rensen, Daniel Risskov and Billinge, Simon JL and Selvan, Raghavendra and others},
  journal={npj Computational Materials},
  volume={8},
  number={1},
  pages={213},
  year={2022},
  publisher={Nature Publishing Group UK London}
}

@article{witman2023defect,
  title={Defect graph neural networks for materials discovery in high-temperature clean-energy applications},
  author={Witman, Matthew D and Goyal, Anuj and Ogitsu, Tadashi and McDaniel, Anthony H and Lany, Stephan},
  journal={Nature Computational Science},
  volume={3},
  number={8},
  pages={675--686},
  year={2023},
  publisher={Nature Publishing Group US New York}
}

@article{davariashtiyani2023formation,
  title={Formation energy prediction of crystalline compounds using deep convolutional network learning on voxel image representation},
  author={Davariashtiyani, Ali and Kadkhodaei, Sara},
  journal={Communications Materials},
  volume={4},
  number={1},
  pages={105},
  year={2023},
  publisher={Nature Publishing Group UK London}
}

@misc{wetzel2025interpretable,
  author       = {{Wetzel, S.J.; Ha, S.; Iten, R.; Klopotek, M.; Liu, Z.}},
  title        = {{Interpretable Machine Learning in Physics: A Review}},
  howpublished = {arXiv. 2025-03. \url{https://arxiv.org/abs/2503.23616}},
  note         = {(accessed 2026-03-19)}
}

@article{luo2024deep,
  title={Deep learning generative model for crystal structure prediction},
  author={Luo, Xiaoshan and Wang, Zhenyu and Gao, Pengyue and Lv, Jian and Wang, Yanchao and Chen, Changfeng and Ma, Yanming},
  journal={npj Computational Materials},
  volume={10},
  number={1},
  pages={254},
  year={2024},
  publisher={Nature Publishing Group UK London}
}

@article{merchant2023scaling,
  title={Scaling deep learning for materials discovery},
  author={Merchant, Amil and Batzner, Simon and Schoenholz, Samuel S and Aykol, Muratahan and Cheon, Gowoon and Cubuk, Ekin Dogus},
  journal={Nature},
  volume={624},
  number={7990},
  pages={80--85},
  year={2023},
  publisher={Nature Publishing Group UK London}
}

@article{allen2022machine,
  title={Machine learning of material properties: Predictive and interpretable multilinear models},
  author={Allen, Alice EA and Tkatchenko, Alexandre},
  journal={Science advances},
  volume={8},
  number={18},
  pages={eabm7185},
  year={2022},
  publisher={American Association for the Advancement of Science}
}

@article{del2023deep,
  title={A deep learning framework to emulate density functional theory},
  author={del Rio, Beatriz G and Phan, Brandon and Ramprasad, Rampi},
  journal={npj Computational Materials},
  volume={9},
  number={1},
  pages={158},
  year={2023},
  publisher={Nature Publishing Group UK London}
}

@article{hegde2023quantifying,
  title={Quantifying uncertainty in high-throughput density functional theory: A comparison of AFLOW, Materials Project, and OQMD},
  author={Hegde, Vinay I and Borg, Christopher KH and Del Rosario, Zachary and Kim, Yoolhee and Hutchinson, Maxwell and Antono, Erin and Ling, Julia and Saxe, Paul and Saal, James E and Meredig, Bryce},
  journal={Physical Review Materials},
  volume={7},
  number={5},
  pages={053805},
  year={2023},
  publisher={APS}
}

@article{nakata2022large,
  title={Large-scale DFT methods for calculations of materials with complex structures},
  author={Nakata, Ayako and Bowler, David R and Miyazaki, Tsuyoshi},
  journal={Journal of the Physical Society of Japan},
  volume={91},
  number={9},
  pages={091011},
  year={2022},
  publisher={The Physical Society of Japan}
}

@article{mao2024transfer,
  title={Transfer learning relaxation, electronic structure and continuum model for twisted bilayer MoTe2},
  author={Mao, Ning and Xu, Cheng and Li, Jiangxu and Bao, Ting and Liu, Peitao and Xu, Yong and Felser, Claudia and Fu, Liang and Zhang, Yang},
  journal={Communications Physics},
  volume={7},
  number={1},
  pages={262},
  year={2024},
  publisher={Nature Publishing Group UK London}
}

@misc{bao2025transfer,
  author       = {{Bao, T.; Mao, N.; Duan, W.; Xu, Y.; Del Maestro, A.; Zhang, Y.}},
  title        = {{Transfer learning electronic structure: millielectron volt accuracy for sub-million-atom moir\'e semiconductor}},
  howpublished = {arXiv. 2025-01. \url{https://arxiv.org/abs/2501.12452}},
  note         = {(accessed 2026-03-19)}
}

@article{digdaya2017interfacial,
  title={Interfacial engineering of metal-insulator-semiconductor junctions for efficient and stable photoelectrochemical water oxidation},
  author={Digdaya, Ibadillah A and Adhyaksa, Gede WP and Trze{\'s}niewski, Bartek J and Garnett, Erik C and Smith, Wilson A},
  journal={Nature communications},
  volume={8},
  number={1},
  pages={15968},
  year={2017},
  publisher={Nature Publishing Group UK London}
}

@article{schindler2024discovery,
  title={Discovery of Stable Surfaces with Extreme Work Functions by High-Throughput Density Functional Theory and Machine Learning},
  author={Schindler, Peter and Antoniuk, Evan R and Cheon, Gowoon and Zhu, Yanbing and Reed, Evan J},
  journal={Advanced Functional Materials},
  volume={34},
  number={19},
  pages={2401764},
  year={2024},
  publisher={Wiley Online Library}
}

@article{lin2023work,
  title={Work function: Fundamentals, measurement, calculation, engineering, and applications},
  author={Lin, Lin and Jacobs, Ryan and Ma, Tianyu and Chen, Dongzheng and Booske, John and Morgan, Dane},
  journal={Physical Review Applied},
  volume={19},
  number={3},
  pages={037001},
  year={2023},
  publisher={APS}
}

@article{weston2018machine,
  title={Machine learning the band gap properties of kesterite I 2-II-IV-V 4 quaternary compounds for photovoltaics applications},
  author={Weston, L and Stampfl, C},
  journal={Physical Review Materials},
  volume={2},
  number={8},
  pages={085407},
  year={2018},
  publisher={APS}
}

@article{nassiri2021high,
  title={High-specific-power flexible transition metal dichalcogenide solar cells},
  author={Nassiri Nazif, Koosha and Daus, Alwin and Hong, Jiho and Lee, Nayeun and Vaziri, Sam and Kumar, Aravindh and Nitta, Frederick and Chen, Michelle E and Kananian, Siavash and Islam, Raisul and others},
  journal={nature Communications},
  volume={12},
  number={1},
  pages={7034},
  year={2021},
  publisher={Nature Publishing Group UK London}
}

@article{bartel2020critical,
  title={A critical examination of compound stability predictions from machine-learned formation energies},
  author={Bartel, Christopher J and Trewartha, Amalie and Wang, Qi and Dunn, Alexander and Jain, Anubhav and Ceder, Gerbrand},
  journal={npj computational materials},
  volume={6},
  number={1},
  pages={97},
  year={2020},
  publisher={Nature Publishing Group UK London}
}

@misc{tian2022information,
  author       = {{Tian, S.I.P.; Walsh, A.; Ren, Z.; Li, Q.; Buonassisi, T.}},
  title        = {{What Information is Necessary and Sufficient to Predict Materials Properties using Machine Learning?}},
  howpublished = {arXiv. 2022-06. \url{https://arxiv.org/abs/2206.04968}},
  note         = {(accessed 2026-03-19)}
}

@article{antunes2022distributed,
  title={Distributed representations of atoms and materials for machine learning},
  author={Antunes, Luis M and Grau-Crespo, Ricardo and Butler, Keith T},
  journal={npj Computational Materials},
  volume={8},
  number={1},
  pages={44},
  year={2022},
  publisher={Nature Publishing Group UK London}
}

@article{peterson2021materials,
  title={Materials discovery through machine learning formation energy},
  author={Peterson, Gordon GC and Brgoch, Jakoah},
  journal={Journal of Physics: Energy},
  volume={3},
  number={2},
  pages={022002},
  year={2021},
  publisher={IOP Publishing}
}

@article{ward2016general,
  title={A general-purpose machine learning framework for predicting properties of inorganic materials},
  author={Ward, Logan and Agrawal, Ankit and Choudhary, Alok and Wolverton, Christopher},
  journal={npj Computational Materials},
  volume={2},
  number={1},
  pages={1--7},
  year={2016},
  publisher={Nature Publishing Group}
}

@article{isayev2017universal,
  title={Universal fragment descriptors for predicting properties of inorganic crystals},
  author={Isayev, Olexandr and Oses, Corey and Toher, Cormac and Gossett, Eric and Curtarolo, Stefano and Tropsha, Alexander},
  journal={Nature communications},
  volume={8},
  number={1},
  pages={15679},
  year={2017},
  publisher={Nature Publishing Group UK London}
}

@article{damewood2023representations,
  title={Representations of materials for machine learning},
  author={Damewood, James and Karaguesian, Jessica and Lunger, Jaclyn R and Tan, Aik Rui and Xie, Mingrou and Peng, Jiayu and G{\'o}mez-Bombarelli, Rafael},
  journal={Annual Review of Materials Research},
  volume={53},
  number={1},
  pages={399--426},
  year={2023},
  publisher={Annual Reviews}
}

@misc{huang2024material,
  author       = {{Huang, C.; Chen, C.; Shi, L.; Chen, C.}},
  title        = {{Material Property Prediction with Element Attribute Knowledge Graphs and Multimodal Representation Learning}},
  howpublished = {arXiv. 2024-11. \url{https://arxiv.org/abs/2411.08414}},
  note         = {(accessed 2026-03-19)}
}

@misc{sanyal2018mt,
  author       = {{Sanyal, S.; Balachandran, J.; Yadati, N.; Kumar, A.; Rajagopalan, P.; Sanyal, S.; Talukdar, P.}},
  title        = {{MT-CGCNN: Integrating Crystal Graph Convolutional Neural Network with Multitask Learning for Material Property Prediction}},
  howpublished = {arXiv. 2018-11. \url{https://arxiv.org/abs/1811.05660}},
  note         = {(accessed 2026-03-19)}
}

@article{unke2019physnet,
  title={PhysNet: A neural network for predicting energies, forces, dipole moments, and partial charges},
  author={Unke, Oliver T and Meuwly, Markus},
  journal={Journal of chemical theory and computation},
  volume={15},
  number={6},
  pages={3678--3693},
  year={2019},
  publisher={ACS Publications}
}

@article{roy2023predicting,
  title={Predicting the work function of 2D MXenes using machine-learning methods},
  author={Roy, Pranav and Rekhi, Lavie and Koh, See Wee and Li, Hong and Choksi, Tej S},
  journal={Journal of Physics: Energy},
  volume={5},
  number={3},
  pages={034005},
  year={2023},
  publisher={IOP Publishing}
}

@article{guo2024machine,
  title={Machine learning facilitated by microscopic features for discovery of novel magnetic double perovskites},
  author={Guo, Shuping and Morrow, Ryan and van den Brink, Jeroen and Janson, Oleg},
  journal={Journal of Materials Chemistry A},
  volume={12},
  number={10},
  pages={6103--6111},
  year={2024},
  publisher={Royal Society of Chemistry}
}

@inproceedings{wong2019netscore,
  title={Netscore: towards universal metrics for large-scale performance analysis of deep neural networks for practical on-device edge usage},
  author={Wong, Alexander},
  booktitle={International Conference on Image Analysis and Recognition},
  pages={15--26},
  year={2019},
  organization={Springer}
}

@misc{kaplan2020scaling,
  author       = {{Kaplan, J.; McCandlish, S.; Henighan, T.; Brown, T.B.; Chess, B.; Child, R.; Gray, S.; Radford, A.; Wu, J.; Amodei, D.}},
  title        = {{Scaling Laws for Neural Language Models}},
  howpublished = {arXiv. 2020-01. \url{https://arxiv.org/abs/2001.08361}},
  note         = {(accessed 2026-03-19)}
}

\clearpage
\pagebreak
\onecolumngrid
\begin{center}
    \textbf{\large TOC Graphic}
\end{center}
\begin{figure}[h]
    \centering
    \includegraphics[width=1\linewidth]{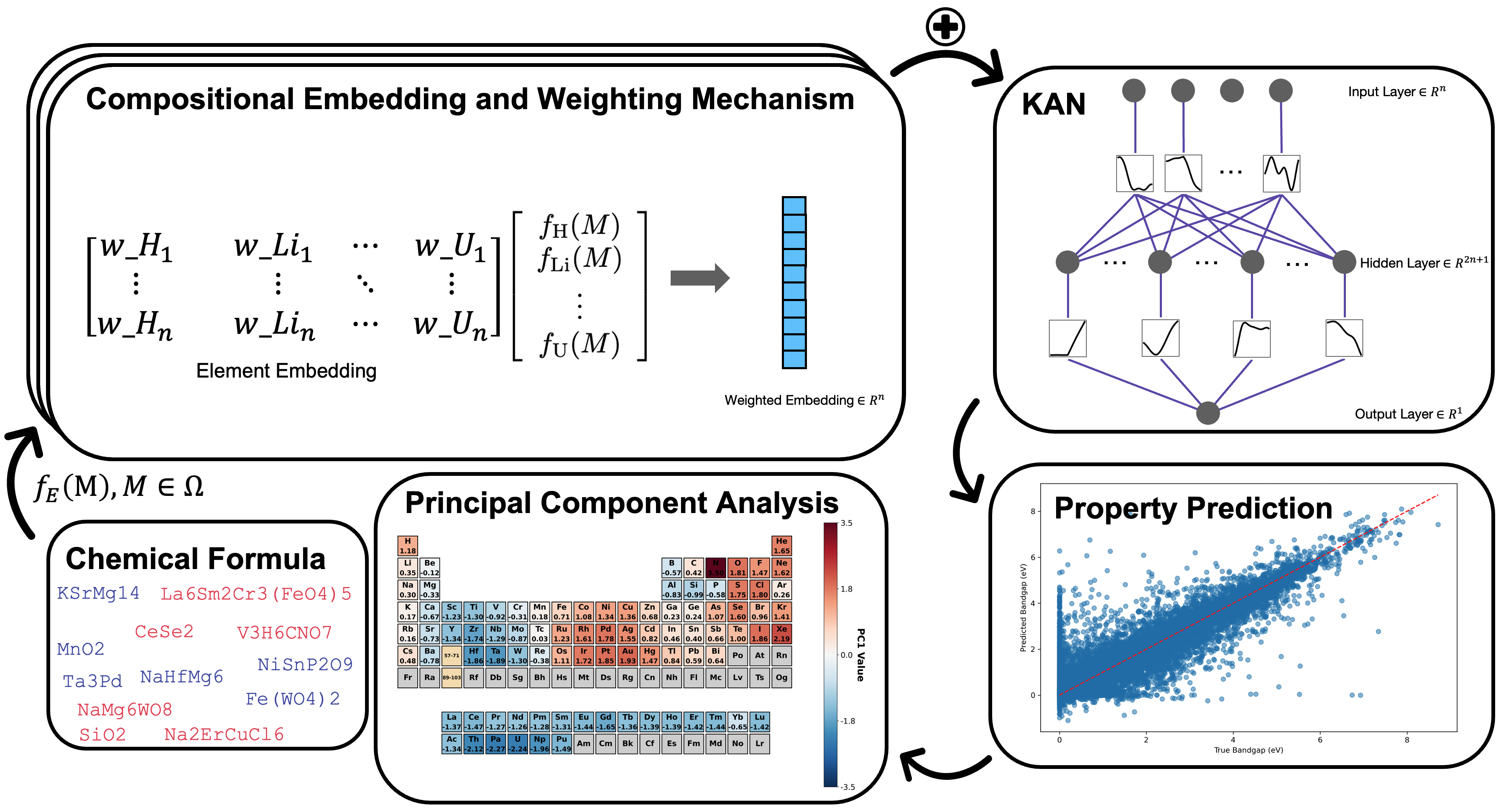}
    \label{fig:TOC Graphic}
\end{figure}

\clearpage
\pagebreak
\onecolumngrid
\begin{center}
\textbf{\large Supplemental Materials}
\end{center}
\setcounter{figure}{0}
\renewcommand{\thefigure}{S\arabic{figure}}
\setcounter{equation}{0}
\renewcommand{\theequation}{S\arabic{equation}}
\setcounter{table}{0}
\renewcommand{\thetable}{S\arabic{table}}

\section*{Details of Data Preparation}
\begin{figure}[h]
    \centering
    \includegraphics[width=\textwidth]{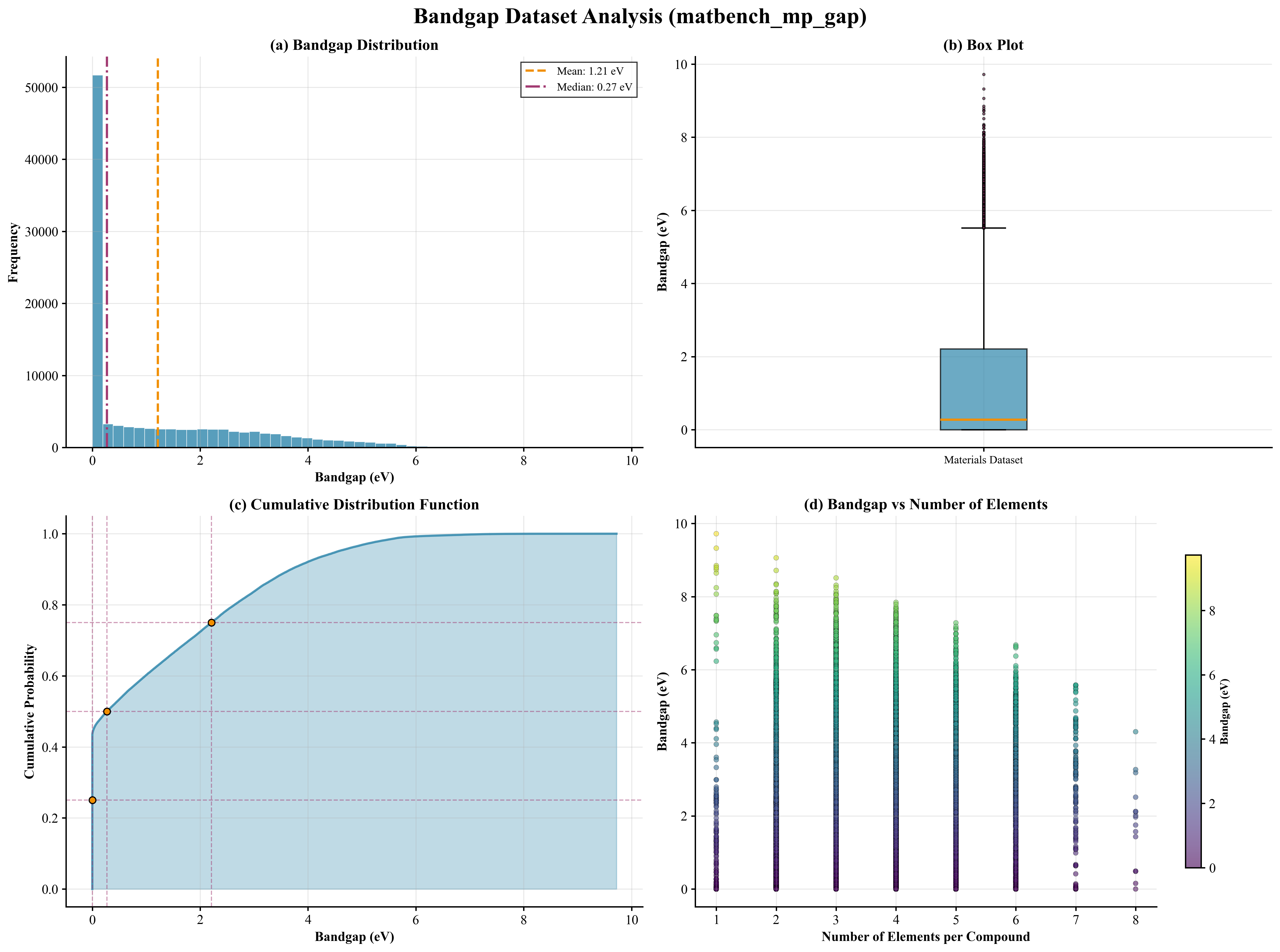}
    \caption{Bandgap dataset analysis (matbench\_mp\_gap). (a) Frequency distribution of bandgap values showing a heavily right-skewed distribution with mean 1.21 eV and median 0.27 eV, indicating a predominance of small-bandgap materials. (b) Box plot summary revealing the statistical spread of bandgap values across the materials dataset, with outliers extending to approximately 10 eV. (c) Cumulative distribution function demonstrating that approximately 25\%, 50\%, and 75\% of materials exhibit bandgaps below 0.27 eV, 1.21 eV, and 2.5 eV, respectively. (d) Scatter plot of bandgap versus number of elements per compound, color-coded by bandgap magnitude, showing that binary and ternary compounds span the full range of bandgap values, while higher-order compounds tend toward smaller bandgaps. The dataset comprises diverse inorganic compounds with bandgaps ranging from 0 to 9.8 eV, providing a comprehensive benchmark for evaluating machine learning models on semiconductor property prediction}
    \label{fig:data_bandgap}
\end{figure}

\begin{figure}[h]
    \centering
    \includegraphics[width=\textwidth]{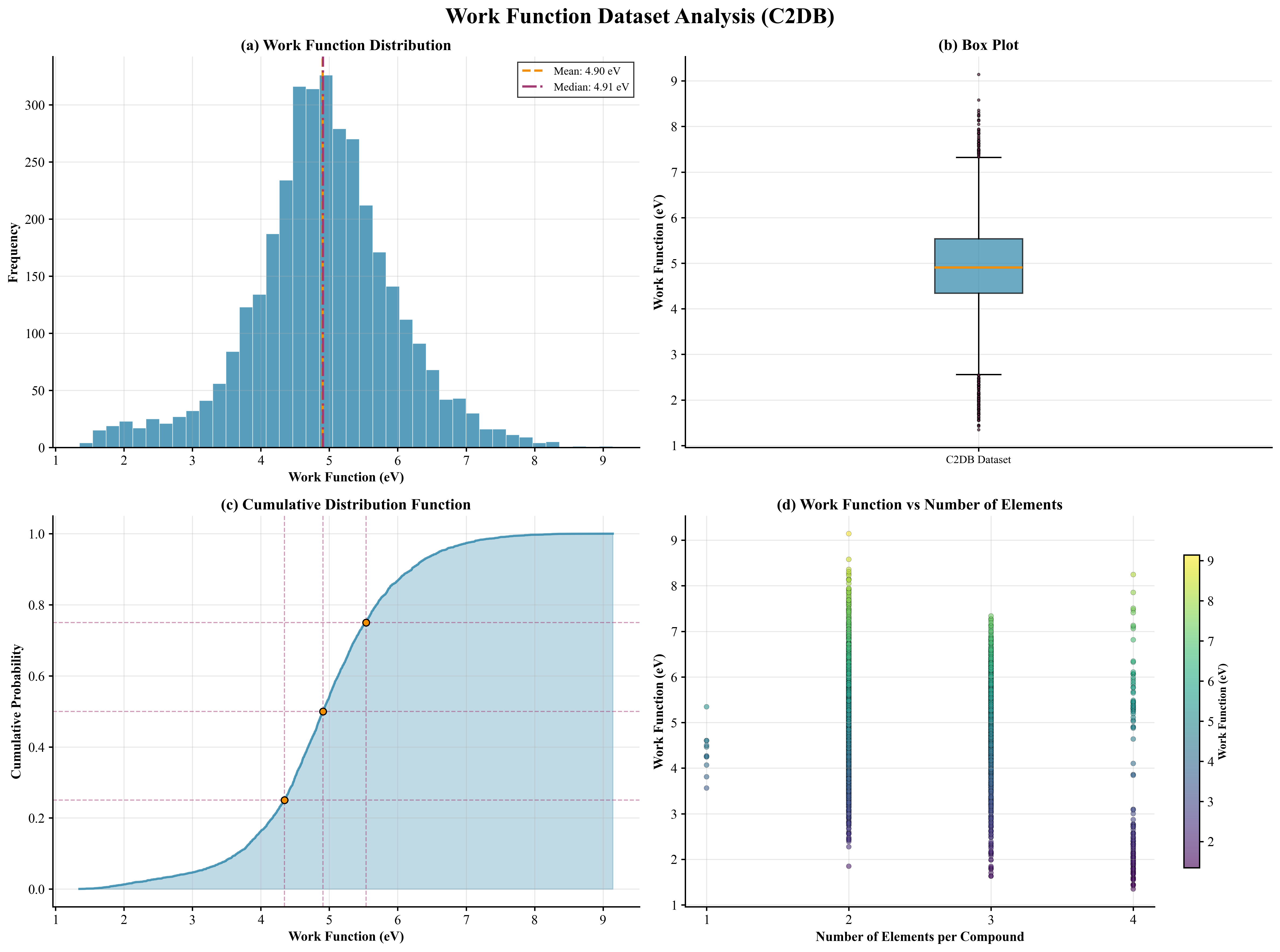}
    \caption{Work function dataset analysis (C2DB). (a) Frequency distribution of work function values exhibiting a near-normal distribution with mean 4.90 eV and median 4.91 eV, centered around typical values for two-dimensional materials. (b) Box plot summary illustrating the statistical spread of work function values across the C2DB dataset, with interquartile range spanning approximately 4.5–5.7 eV and outliers extending to both lower and higher extremes. (c) Cumulative distribution function showing that 25\%, 50\%, and 75\% of materials exhibit work functions below approximately 4.5 eV, 4.9 eV, and 5.7 eV, respectively, with vertical dashed lines marking key quartiles. (d) Scatter plot of work function versus number of elements per compound, color-coded by work function magnitude, revealing that the dataset is dominated by binary compounds with relatively uniform work function distribution, while ternary and quaternary systems show similar ranges but with reduced sample density. The C2DB dataset provides a comprehensive collection of two-dimensional materials with work functions ranging from approximately 1.5 to 9.0 eV, representing diverse electronic properties relevant for surface science and device applications}
    \label{fig:data_workfunction}
\end{figure}

\begin{figure}[h]
    \centering
    \includegraphics[width=\textwidth]{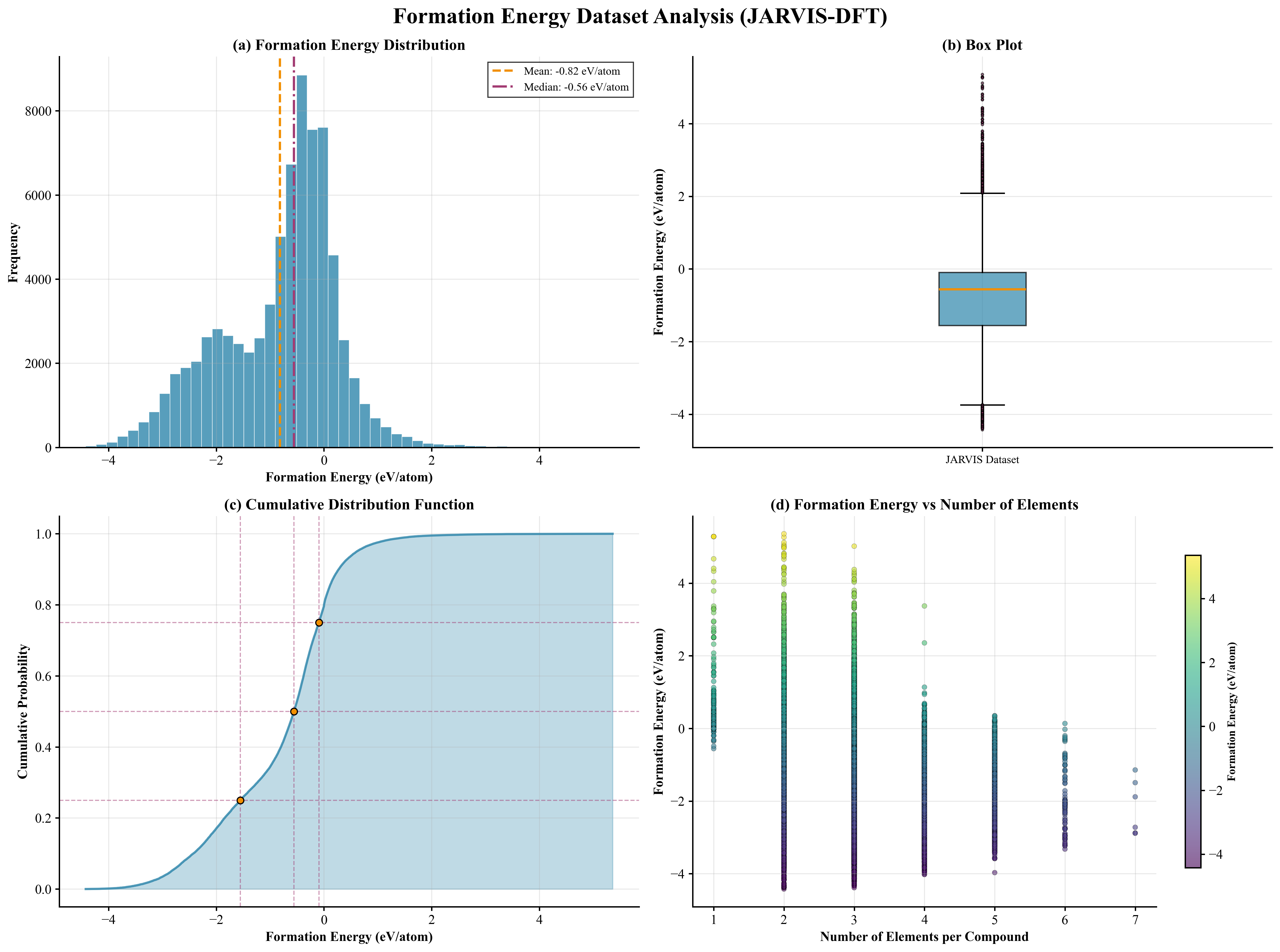}
    \caption{Formation energy dataset analysis (JARVIS-DFT). (a) Frequency distribution of formation energies showing a slightly right-skewed distribution with mean $-0.82$ eV/atom and median $-0.56$ eV/atom, indicating that most compounds in the dataset are thermodynamically stable relative to their constituent elements. (b) Box plot summary revealing the statistical spread of formation energies, with the majority of compounds exhibiting negative formation energies (thermodynamically favorable) and outliers extending to both highly stable ($-4.5$ eV/atom) and metastable (5.0 eV/atom) regimes. (c) Cumulative distribution function demonstrating that approximately 25\%, 50\%, and 75\% of materials exhibit formation energies below $-1.5$ eV/atom, $-0.56$ eV/atom, and 0 eV/atom, respectively. (d) Scatter plot of formation energy versus number of elements per compound, color-coded by formation energy magnitude, showing that binary and ternary compounds dominate the dataset and span the full stability range, while higher-order compounds tend toward intermediate formation energies. The JARVIS-DFT dataset encompasses diverse inorganic compounds with formation energies ranging from $-4.5$ to 5.0 eV/atom, providing a comprehensive benchmark for thermodynamic stability prediction across different chemical compositions and structural complexities}
    \label{fig:data_formationenergy}
\end{figure}

The datasets utilized in this study were sourced from established materials databases to ensure reproducibility and comparability with existing benchmarks. Three distinct property prediction tasks were examined: formation energy prediction using the JARVIS-DFT dataset, bandgap prediction using the Materials Project dataset (matbench\_mp\_gap), and work function prediction using the C2DB dataset. The statistical distributions and compositional characteristics of these datasets are summarized in Figures~\ref{fig:data_bandgap}-\ref{fig:data_formationenergy}, revealing diverse property ranges and chemical compositions that provide comprehensive benchmarks for model evaluation.

A critical preprocessing step involves converting chemical formulas into numerical representations suitable for machine learning models. Each chemical formula was parsed using the pymatgen Composition class to extract elemental constituents and their stoichiometric coefficients. The raw elemental compositions were then normalized by the total number of atoms to obtain fractional atomic compositions, ensuring scale invariance across compounds with different formula unit sizes.

Mathematically, for a compound with chemical formula $A_a B_b C_c \ldots$, the normalized element vector is defined as:
$$\mathbf{v}_i = \frac{n_i}{\sum_j n_j}$$
where $\mathbf{v}_i$ represents the fractional composition of element $i$, and $n_i$ denotes the stoichiometric coefficient of element $i$ in the chemical formula. This normalization approach ensures that compounds such as $\mathrm{SiO_2}$ and $\mathrm{Si_2O_4}$ yield identical representations, focusing the model on compositional ratios rather than absolute quantities.

The complete element vocabulary was constructed by identifying all unique elements present across the entire dataset. Elements were then mapped to consistent integer indices, creating a standardized representation scheme. For the formation energy task, this procedure yielded a vocabulary of 89 elements spanning the periodic table from hydrogen to uranium, reflecting the diverse chemical space covered by the JARVIS-DFT database.

Data partitioning followed a stratified random sampling approach to ensure representative distributions across training and validation sets. The datasets were split using an 80:20 ratio for training and testing, respectively, with a fixed random seed (42) to ensure reproducibility. This splitting strategy maintains statistical consistency between training and validation sets while providing sufficient data for both model optimization and unbiased performance evaluation.

Prior to model training, data quality assurance procedures were implemented to identify and remove problematic entries. Compounds with malformed chemical formulas or missing property values were excluded from the analysis. Additionally, extreme outliers beyond three standard deviations from the mean were flagged for manual inspection to prevent potential data entry errors from affecting model training.

The final input representation combines the normalized elemental compositions with learnable element embeddings within the KAN architecture. Unlike traditional approaches that rely on extensive feature engineering or pre-computed descriptors, this methodology allows the model to learn task-specific elemental representations directly from the training data. The embedding dimension was systematically varied (3, 5, 7, and 10 dimensions) to investigate the trade-off between representational capacity and model complexity.

Data preprocessing also included target property scaling where appropriate. For formation energies, values were retained in their original units (eV/atom) to maintain physical interpretability. Bandgap and work function values were similarly preserved in electronvolts to facilitate direct comparison with experimental measurements and literature values.

This comprehensive data preparation pipeline ensures robust model training while preserving the chemical and physical meaning of the input features, enabling meaningful interpretation of the learned representations and their correlation with fundamental atomic properties.

\section*{Benchmark for ReLU}
\begin{figure}[h]
    \centering
    \includegraphics[width=\textwidth]{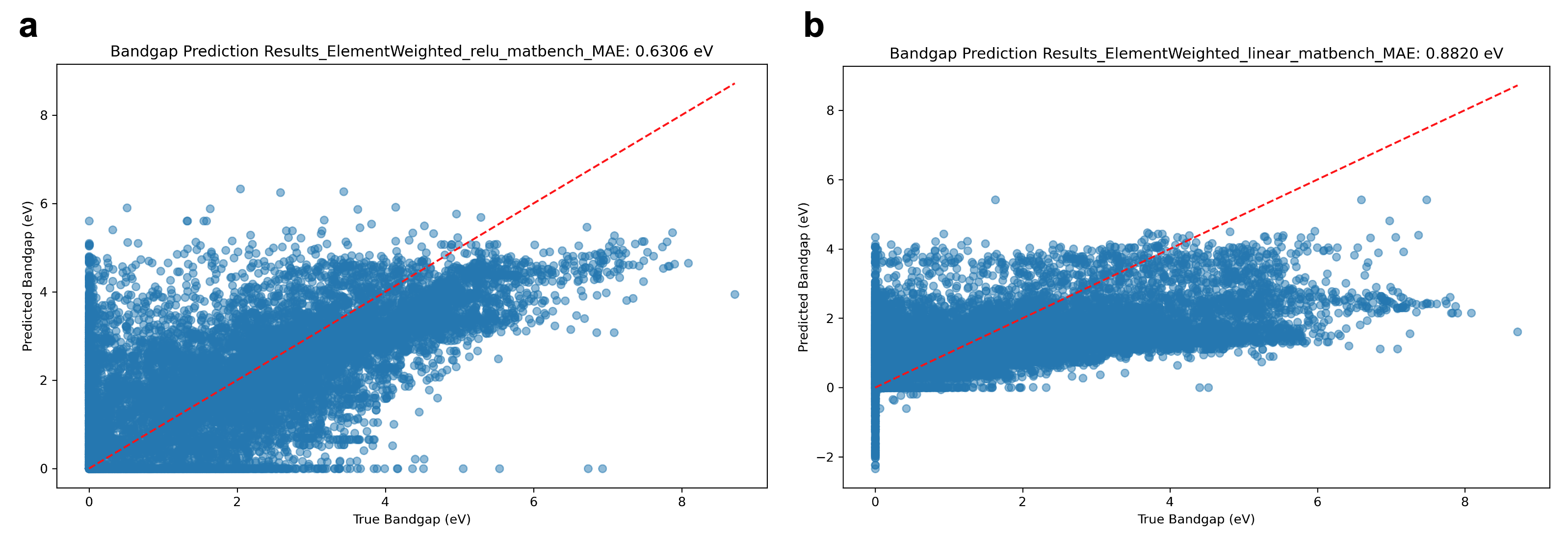}
    \caption{Bandgap prediction for the ElementWeighted Model. (a) use the $\text{ReLU}(\cdot)$ as the activation function to strict the final output. (b) use the linear combination of the weighted contribution for each element as the output directly}
    \label{fig:EW_matbench}
\end{figure}

According to A. Ma's paper\cite{ma2025predicting}, a single test of the bandgap prediction results with linear combination or $\text{ReLU}(\cdot)$ activation function is shown in Figure~\ref{fig:EW_matbench}. The detailed MAE results with the same approach as in their paper are shown in Table~\ref{tab: EW_mae}. The entire dataset is only divided into the training dataset and the validation dataset by 8:2 and the validation dataset is used to replace the test dataset to make the final prediction but with the 10-fold cross-validation to decrease the effect of data leakage problem in this case.

\begin{table}[h]
\caption{MAE results comparison for $\text{ReLU}(\cdot)$ and linear combination method under 10-fold cross-validation approach}
\centering
\begin{tabular}{lllll}
\cline{1-2}
\textbf{Model} & \textbf{Mean Absolute Error (eV)} &  &  &  \\ \cline{1-2}
Linear Model, $\varepsilon_{\text{linear}}(\cdot)$   & 0.8835±0.0071                     &  &  &  \\ \cline{1-2}
ReLU Model, $\varepsilon_{\text{relu}}(\cdot)$     & 0.6344±0.0071                     &  &  &  \\ \cline{1-2}
               &                                   &  &  & 
\end{tabular}
\label{tab: EW_mae}
\end{table}

\begin{figure}[h]
    \centering
    \noindent\includegraphics[width=0.9\textwidth]{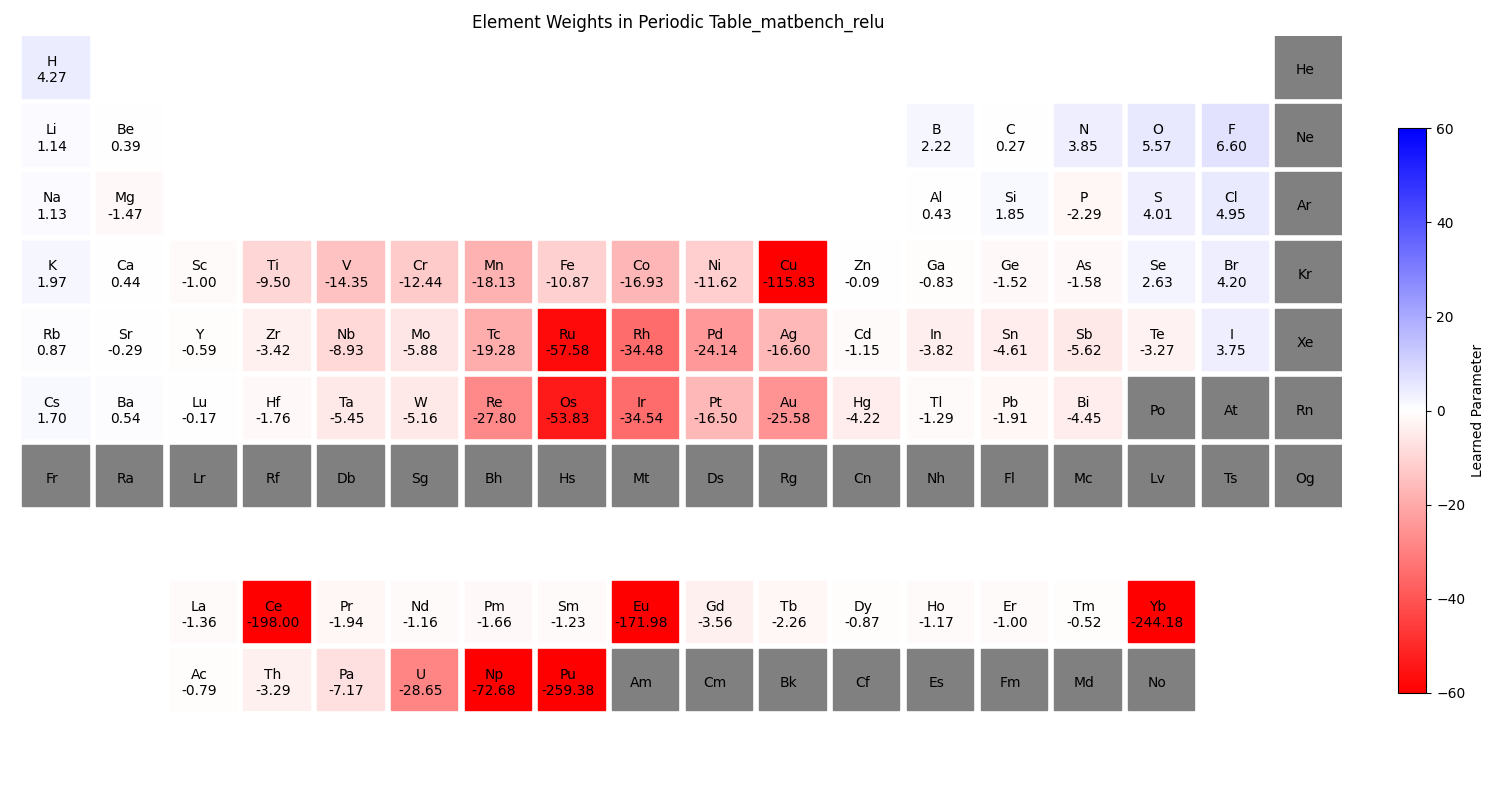}
    \caption{Periodic table visualization of the learned parameters}
    \label{fig:periodic_table_relu}
\end{figure}

Furthermore, the visualization results of the learned parameters for each element in the periodic table format for these two different integration approaches, $\text{ReLU}(\cdot)$ and linear combination, are shown here in Figure~\ref{fig:periodic_table_relu}, as well.

\begin{figure}[h]
    \centering
    \noindent\includegraphics[width=\textwidth]{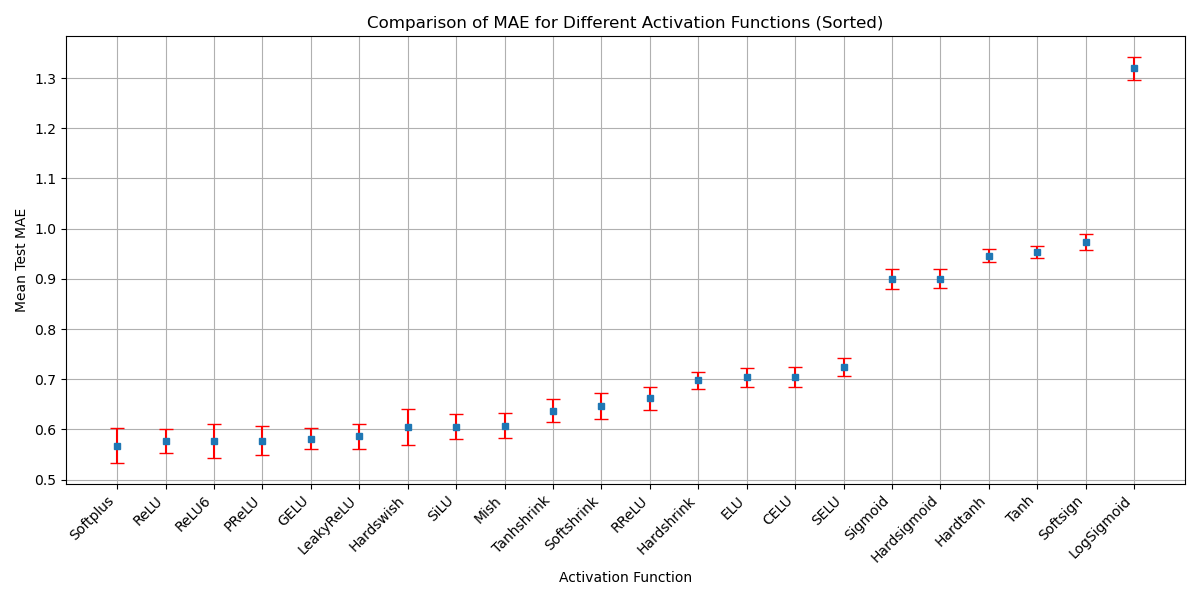}
    \caption{Different pre-defined activation function test for ElementWeighted model}
    \label{fig:activation_function}
\end{figure}

In order to understand the way they handle this problem, a supplement test of different types of pre-defined activation functions in PyTorch package is done. The result in Figure~\ref{fig:activation_function} can demonstrate the reasonability of the choice of $\text{ReLU}(\cdot)$ in this specific model and prediction task to some extent.

\clearpage
\section*{Efficiency Score}
\begin{figure}[h]
    \centering
    \includegraphics[width=1\linewidth]{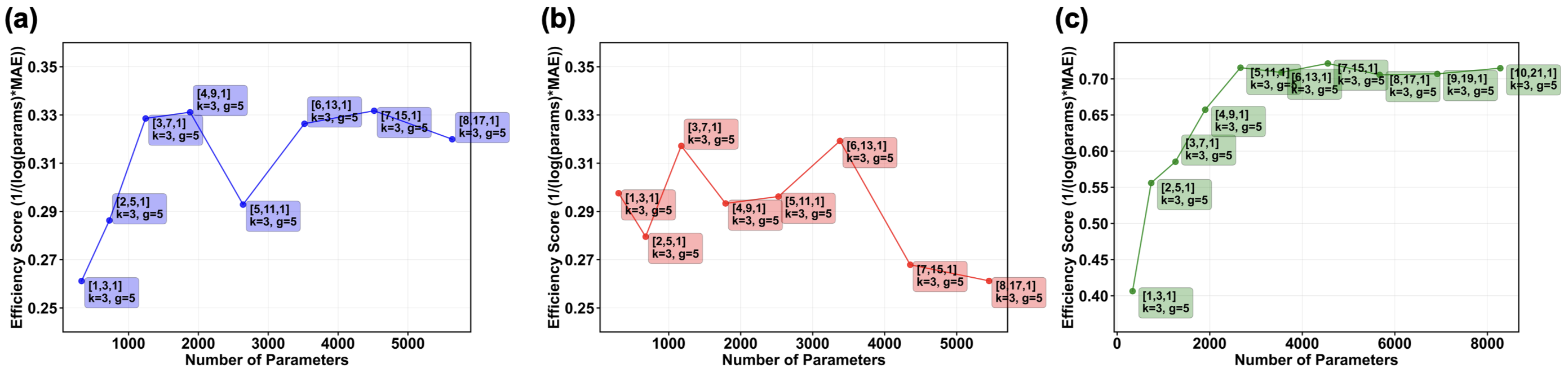}
    \caption{Simple efficiency score under different KAN structure for three different properties}
    \label{fig:efficiency}
\end{figure}

\section*{Activation Function Monitoring}
\subsection{Bandgap}
\begin{figure}[h]
    \centering
    \includegraphics[width=1\linewidth]{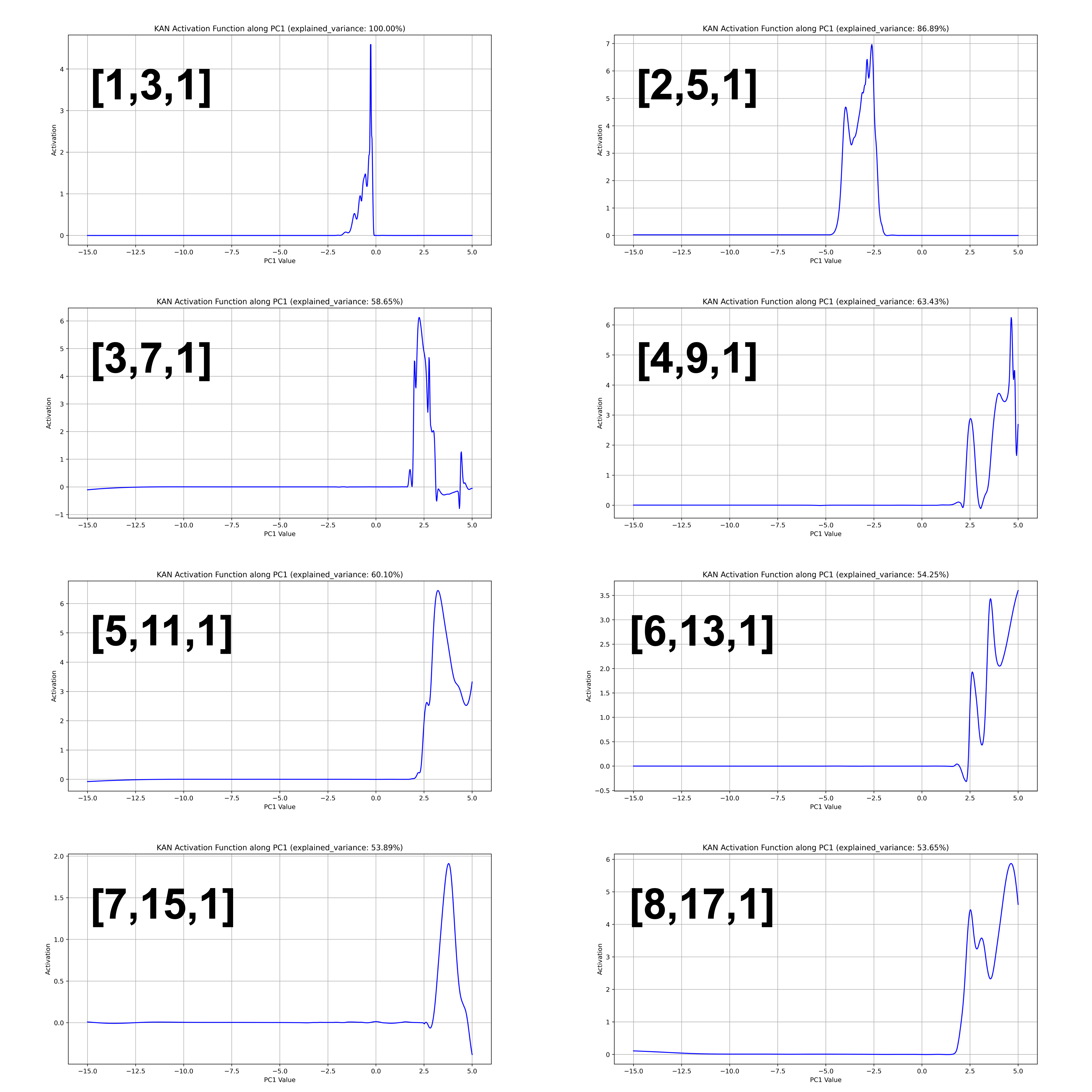}
    \caption{Evolution of KAN activation functions along PC1 across different architectures for bandgap. The explained variance percentage indicates the information concentration in the primary component, revealing the learning convergence and functional complexity evolution from simple structures to the optimal $[7,15,1]$ configuration}
    \label{fig:bandgap_pc1_evolution}
\end{figure}
To understand the physical basis of this singularity in Fig.~\ref{fig:bandgap_accuracy}c, we analyze the evolution of learned activation functions along the first principal component (PC1) across different KAN architectures (Fig.~\ref{fig:bandgap_pc1_evolution}). The PC1 analysis reveals a compelling narrative of learning convergence and functional optimization:

\textbf{Early-stage learning} ($[1,3,1]$ to $[3,7,1]$): The activation functions evolve from overly simplistic single peaks (100\% variance explained) to increasingly complex multi-modal structures (58.65\% variance explained). This phase represents the network's exploration of the feature space, gradually discovering the nonlinear relationships governing bandgap formation.

\textbf{Intermediate complexity} ($[4,9,1]$ to $[6,13,1]$): The explained variance stabilizes around 54-63\%, indicating that the network has identified the intrinsic dimensionality of the bandgap prediction problem. The activation functions during this phase exhibit diverse morphologies, suggesting active learning of distinct physical mechanisms.

\textbf{Optimal convergence} ($[7,15,1]$): At the singularity point, the PC1 activation function achieves an elegant balance—retaining sufficient complexity to capture essential physics while avoiding over-parameterization. The 53.89\% explained variance suggests optimal information distribution across multiple components, neither over-concentrated nor over-dispersed.

\textbf{Saturation regime} ($[8,17,1]$): The minimal change in both explained variance (53.65\%) and activation function morphology indicates learning saturation. The network has essentially learned all extractable information from the data, and additional parameters only introduce redundancy rather than new physical insights.

The physical interpretation of this evolution suggests that the $[7,15,1]$ configuration captures the essential quantum mechanical principles governing electronic band structure. The seven input dimensions likely correspond to fundamental electronic descriptors (s, p, d orbital contributions, electronegativity, etc.), while the fifteen hidden nodes may represent distinct electronic states or hybridization patterns crucial for bandgap determination. The convergence of the activation function to a stable, interpretable form provides compelling evidence that KAN has learned genuine physical relationships rather than mere statistical correlations.

\subsection{Work Function}
The work function prediction results reveal a distinct performance singularity at the $[6,13,1]$ configuration, contrasting with bandgap prediction which reached optimal performance at $[7,15,1]$. This difference reflects the fundamental distinction in physical complexity between surface and bulk electronic properties. Analysis of PC1 explained variance shows convergence from 100\% ([1,3,1]) to a stable 24-25\% range for configurations $[6,13,1]$ through $[8,17,1]$, indicating that the network has discovered the intrinsic dimensionality of work function prediction. The rapid stabilization of both explained variance and activation function morphologies provides compelling evidence for learning saturation at $[6,13,1]$.

Individual dimension activation functions demonstrate clear functional convergence at the optimal architecture, with each dimension developing distinct, non-redundant roles in representing surface electronic structure. The six input dimensions likely correspond to fundamental surface descriptors such as surface atomic coordination, d-orbital occupancy, surface relaxation effects, and interfacial electronic redistribution, while the thirteen hidden nodes may represent distinct surface electronic states critical for work function determination. Further architectural expansion to $[7,15,1]$ and $[8,17,1]$ introduces parameter redundancy without contributing new physical insights, resulting in performance degradation rather than improvement.

This saturation behavior contrasts markedly with bandgap prediction, where bulk electronic band structure requires more complex representations. The $[6,13,1]$ optimum validates the hypothesis that KAN architectures naturally adapt to the intrinsic complexity of different physical phenomena, with surface properties requiring less complex networks than bulk electronic properties. The convergence of multiple metrics—performance plateau, PC1 stability, and morphological consistency—demonstrates that the network has extracted all learnable information about surface electronic structure from the available data, representing a fundamental limit where additional complexity yields diminishing returns.
\begin{figure}[h]
    \centering
    \includegraphics[width=1\linewidth]{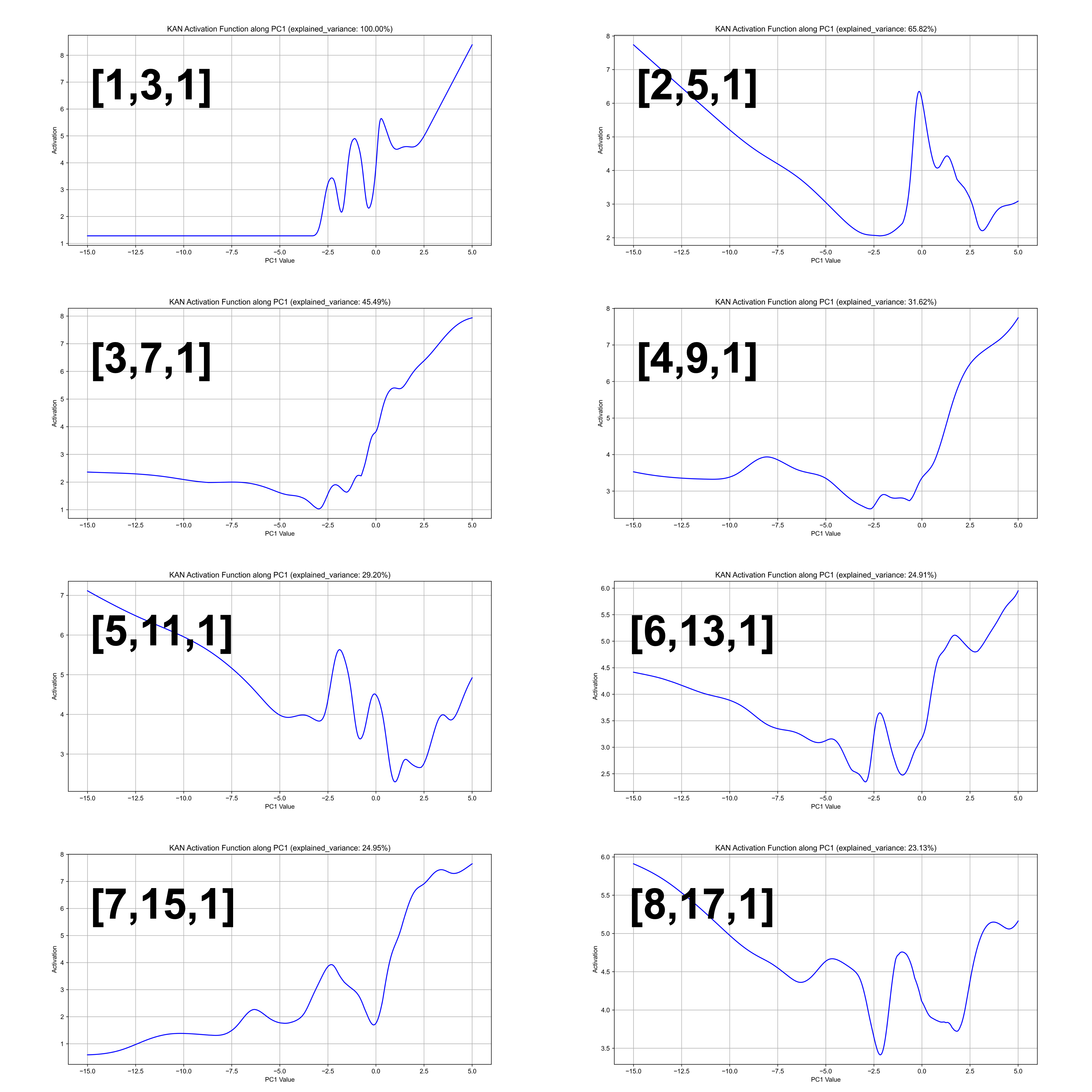}
    \caption{Evolution of KAN activation functions along PC1 across different architectures for work function. The explained variance percentage indicates the information concentration in the primary component, revealing the learning convergence and functional complexity evolution from simple structures to the optimal $[6,13,1]$ configuration}
    \label{fig:workfunction_pc1_evolution}
\end{figure}

\subsection{Formation Energy}
\begin{figure}[h]
    \centering
    \includegraphics[width=1\linewidth]{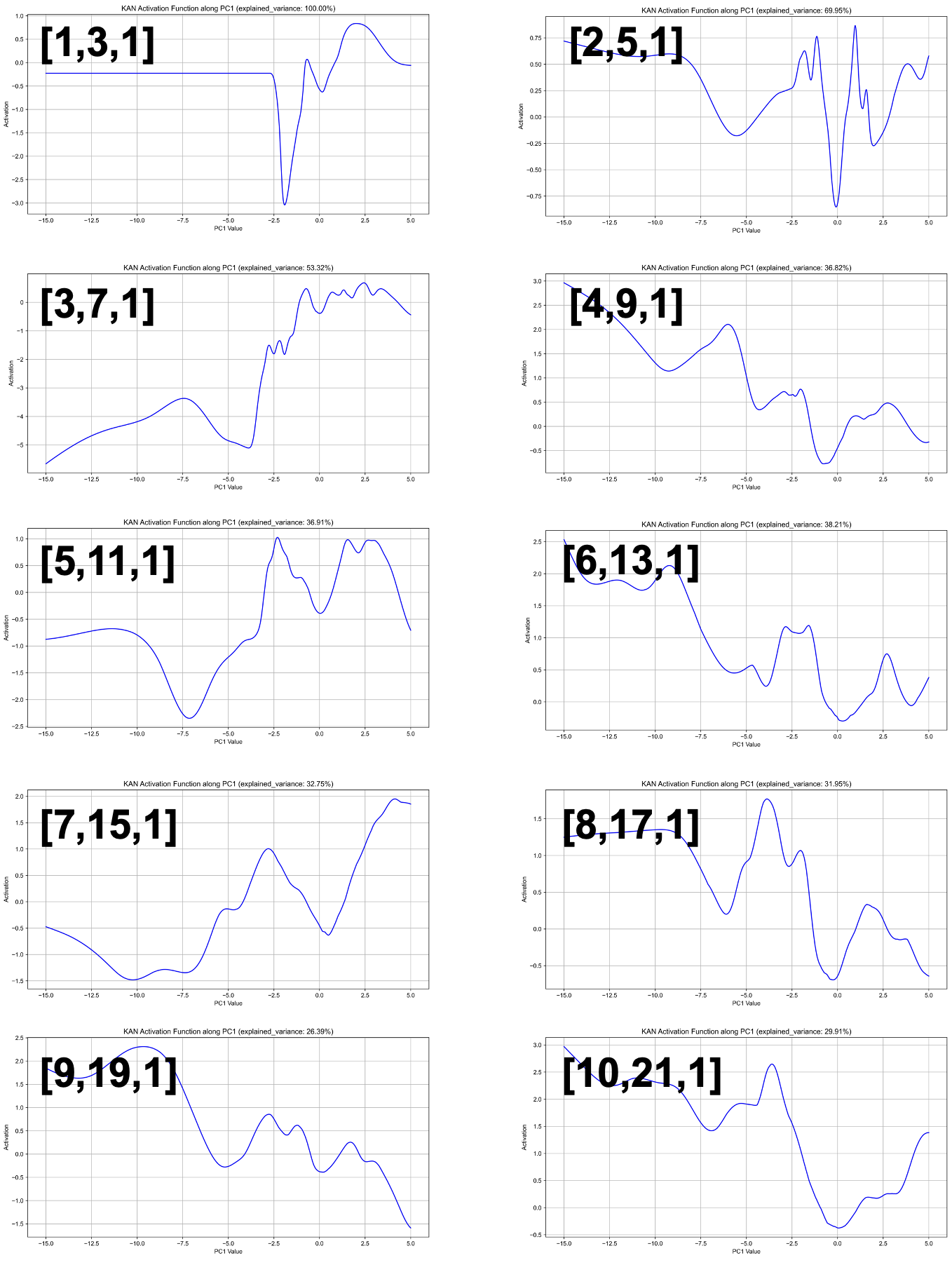}
    \caption{Evolution of KAN activation functions along PC1 across different architectures for formation energy. The explained variance percentage indicates the information concentration in the primary component, revealing the learning convergence and functional complexity evolution from simple structures to the optimal $[10,21,1]$ configuration}
    \label{fig:formationenergy_pc1_evolution}
\end{figure}

Formation energy prediction reveals distinctly different learning dynamics compared to bandgap and work function predictions, providing insights into the relative complexity of different materials properties. Unlike the clear performance singularities and learning saturation observed for electronic properties, formation energy exhibits continuous improvement across all tested architectures from $[1,3,1]$ to $[10,21,1]$, with no evidence of performance plateaus.

The most telling difference lies in the PC1 explained variance evolution. While bandgap and work function show stabilization of explained variance at larger architectures (stabilizing around 54\% and 24\% respectively), formation energy demonstrates continued fluctuation without convergence. The explained variance oscillates between 26.39\% and 38.21\% across the largest tested architectures, with the final $[10,21,1]$ configuration showing 29.91\% explained variance—indicating that the network is still actively discovering new informational dimensions rather than reaching a stable representation.

This behavior reflects the inherent complexity of formation energy, which depends on intricate many-body interactions, crystal structure stability, chemical bonding effects, and compositional dependencies that require more sophisticated mathematical representations than the primarily electronic phenomena governing bandgap and work function. The absence of learning saturation suggests that even larger architectures beyond $[10,21,1]$ might continue to yield performance improvements, representing a fundamental challenge for efficient model design in thermodynamic property prediction.

\clearpage
\section{Principal Component Analysis}
\begin{figure}[h]
    \centering
    \includegraphics[width=\textwidth]{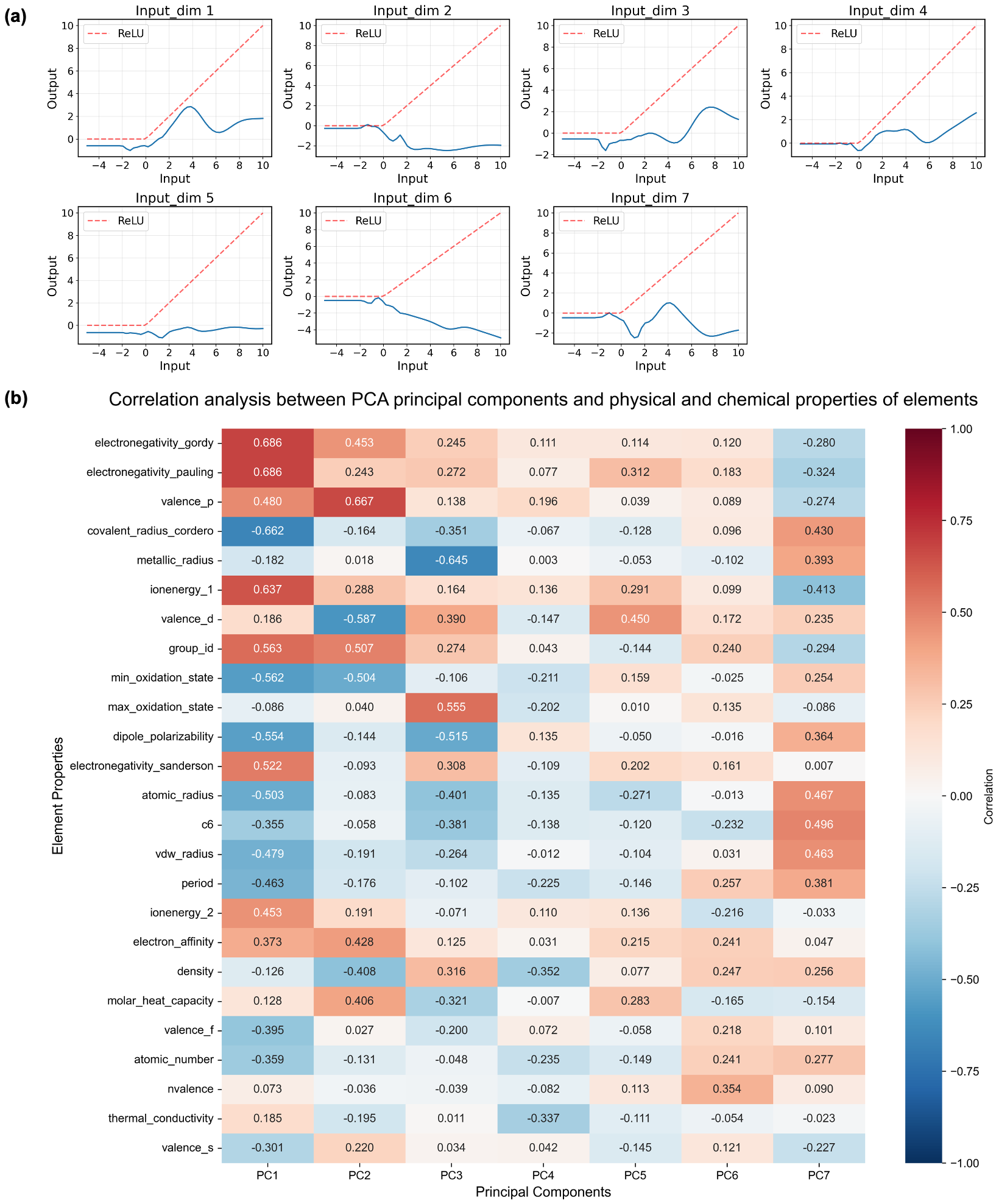}
    \caption{(a) Learned activation functions of the seven input dimensions in the KAN model, compared with the standard ReLU (red dashed line). The learned nonlinearities deviate substantially from the fixed ReLU form, indicating that the model adapts its internal representation to capture chemically relevant transformations of the input features. (b) Correlation matrix between PCA principal components (PC1–PC7) and tabulated elemental properties. PC1 shows the strongest correlations with electronegativity (Pauling and Gordy scales) and covalent radius, confirming that the dominant latent axis learned by the model encodes a chemically meaningful “electronegativity–radius” dimension}
    \label{fig:formationenergy_pca}
\end{figure}

To quantitatively substantiate this connection, we performed correlation analysis between PCA-derived components and a wide range of tabulated elemental properties (Fig.~\ref{fig:formationenergy_pca}). Among all properties considered, PC1 exhibited the strongest positive correlation with electronegativity (Pauling and Gordy scales, $r = 0.686$) and a strong negative correlation with covalent radius ($r = –0.662$). These results confirm that PC1 effectively captures the same physical axis that links high electronegativity to small atomic radius, and vice versa. Consequently, comparing PC1 with normalized Pauling electronegativity provides the most direct and chemically interpretable validation of what the model has learned.

\begin{figure}[h]
    \centering
    \includegraphics[width=0.6\textwidth]{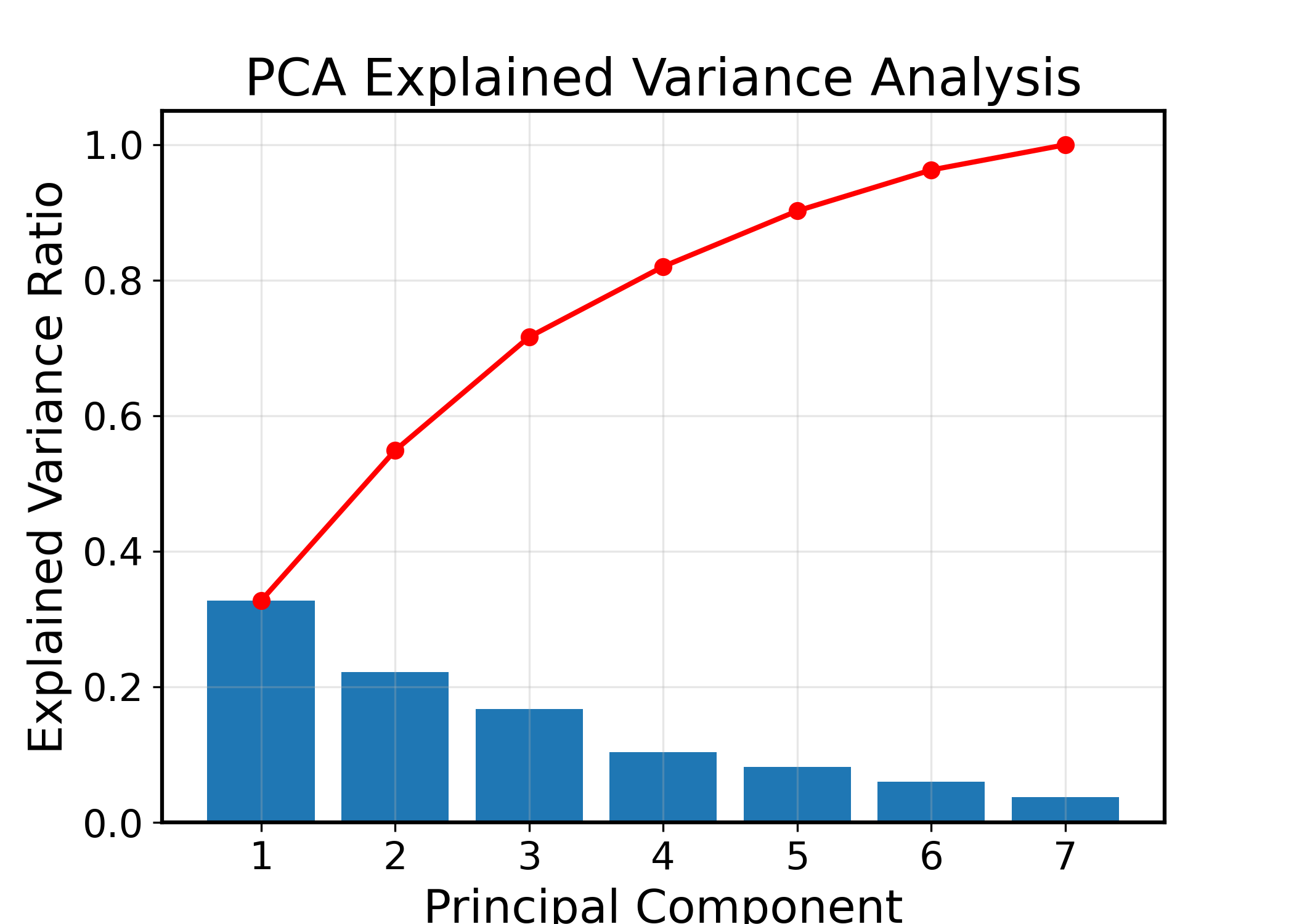}
    \caption{Explained variance ratio of the seven principal components. The bar plot indicates the variance explained by each individual component, while the red line shows the cumulative explained variance. PC1 alone accounts for the largest fraction of the variance, and the first three PCs together explain over 70\% of the total variance, demonstrating that a small number of latent dimensions capture the dominant chemical trends in the input representation}
    \label{fig:pca_var}
\end{figure}

\begin{figure}[h]
    \centering
    \includegraphics[width=\textwidth]{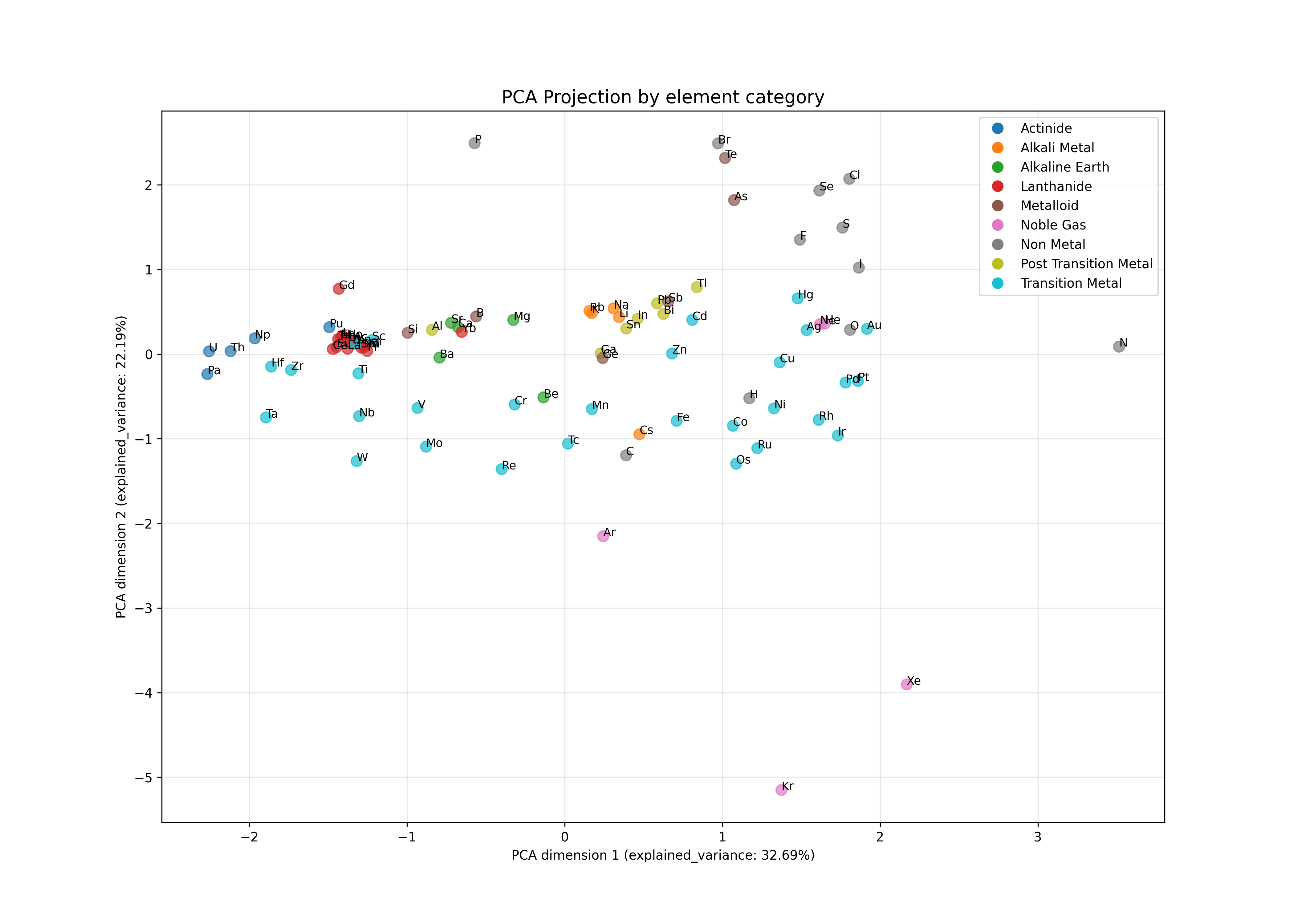}
    \caption{PCA projection of elements colored by chemical category. The first two principal components (PC1 and PC2) capture 32.69\% and 22.19\% of the variance, respectively. Clear grouping of elements according to their chemical families (e.g., alkali metals, transition metals, halogens, noble gases) indicates that the latent space preserves chemically meaningful structure, with separation driven largely by electronegativity and atomic radius trends}    
    \label{fig:pca_pc1_pc2}
\end{figure}

\clearpage

\clearpage
\section*{Benchmark For Complicated Neural Network}

\subsection{Parameter Efficiency Definition}

To provide a comprehensive evaluation of model parameter efficiency, we adopt the NetScore methodology \cite{wong2019netscore} which establishes a power-law relationship between model performance and parameter complexity. This approach addresses the limitation of linear efficiency assumptions and provides a more theoretically sound framework for comparing models across different scales.

The parameter efficiency is defined using a power-law scaling relationship:

\begin{equation}
\text{MAE} = C \times \text{Parameters}^{-\beta/\alpha}
\label{eq:power_law}
\end{equation}

where $C$ is the efficiency constant, $\alpha$ represents the performance importance weight, and $\beta$ represents the parameter complexity penalty.

The efficiency metric is then calculated as:

\begin{equation}
\text{Efficiency} = \frac{(1/\text{MAE})^{\alpha}}{\text{Parameters}^{\beta}}
\label{eq:efficiency}
\end{equation}

Based on established scaling laws in deep learning \cite{kaplan2020scaling}, we set:
\begin{itemize}
    \item $\alpha = 1.5$: Performance importance weight, emphasizing the non-linear significance of accuracy improvements
    \item $\beta = 0.6$: Parameter complexity penalty, reflecting the typical relationship between model size and computational cost
\end{itemize}

This parameterization results in a power-law exponent of $-\beta/\alpha = -0.4$, which aligns with empirical observations that doubling parameters typically reduces error by approximately 25\%.

The efficiency iso-lines in our performance plots represent models with equivalent parameter efficiency under this framework. These lines follow the relationship:

\begin{equation}
\log(\text{MAE}) = \log(C) - \frac{\beta}{\alpha} \log(\text{Parameters})
\label{eq:iso_lines}
\end{equation}

where different values of $C$ correspond to different efficiency levels.

\subsection{Bandgap Prediction Performance}
Table~\ref{tab:bandgap_performance} presents the comparative performance of models on the MatBench MP Gap dataset for the bandgap prediction.

\begin{table}[h]
\centering
\caption{Bandgap Prediction Model Performance (MatBench MP Gap Dataset)}
\label{tab:bandgap_performance}
\begin{tabular}{lccc}
\toprule
\textbf{Model Name} & \textbf{MAE (eV)} & \textbf{Parameters (K)} & \textbf{Efficiency} \\
\midrule
CGCNN & 0.2280 & 128.0 & 0.499772 \\
MEGNet & 0.2350 & 168.0 & 0.405707 \\
ALIGNN & 0.2500 & 4030.0 & 0.054942 \\
\textbf{EWKAN (This Work)} & \textbf{0.3582} & \textbf{4.5} & \textbf{1.891839} \\
\bottomrule
\end{tabular}
\end{table}

\subsection{Formation Energy Prediction Performance}
Table~\ref{tab:formation_energy_performance} shows the comprehensive comparison of models for the formation energy prediction.

\begin{table}[h]
\centering
\caption{Formation Energy Prediction Model Performance (MatBench Discovery Dataset)}
\label{tab:formation_energy_performance}
\scriptsize
\begin{tabular}{lccc}
\toprule
\textbf{Model Name} & \textbf{MAE (eV/atom)} & \textbf{Parameters (K)} & \textbf{Efficiency} \\
\midrule
eSEN-30M-OAM & 0.0180 & 30200.0 & 0.849351 \\
SevenNet-MF-ompa & 0.0210 & 25700.0 & 0.742525 \\
GRACE-2L-OAM & 0.0230 & 12600.0 & 0.993541 \\
DPA-3.1-3M-FT & 0.0230 & 3270.0 & 2.231925 \\
ORB v3 & 0.0240 & 25500.0 & 0.610603 \\
AlphaNet-v1-OMA & 0.0240 & 4650.0 & 1.695163 \\
MatterSim v1 5M & 0.0240 & 4550.0 & 1.717419 \\
MACE-MPA-0 & 0.0280 & 9060.0 & 0.901542 \\
GRACE-1L-OAM & 0.0340 & 3450.0 & 1.202514 \\
GNoME & 0.0350 & 16200.0 & 0.455186 \\
ALIGNN & 0.0930 & 4030.0 & 0.242155 \\
ESNet & 0.1090 & 5430.0 & 0.159581 \\
Wrenformer & 0.1100 & 5170.0 & 0.162113 \\
CGCNN+P & 0.1130 & 128.0 & 1.432374 \\
BOWSR & 0.1180 & 168.0 & 1.140227 \\
MEGNet & 0.1300 & 168.0 & 0.986051 \\
CGCNN & 0.1380 & 128.0 & 1.061342 \\
Voronoi RF & 0.1480 & 26200.0 & 0.039231 \\
\textbf{EWKAN (This Work)} & \textbf{0.1551} & \textbf{8.1} & \textbf{4.666500} \\
\bottomrule
\end{tabular}
\end{table}

The analysis using the NetScore efficiency framework reveals that EWKAN demonstrates exceptional parameter efficiency across both benchmarks:

\begin{itemize}
    \item \textbf{Bandgap Prediction}: EWKAN achieves an efficiency score of 1.891839, which is 3.78× higher than the second-best model (CGCNN: 0.499772)
    \item \textbf{Formation Energy Prediction}: EWKAN reaches an efficiency score of 4.666500, surpassing the second-best model (DPA-3.1-3M-FT: 2.231925) by 2.09×
    \item \textbf{Ultra-compact Architecture}: EWKAN consistently uses orders of magnitude fewer parameters (4.5K-8.1K) compared to conventional models (128K-30,200K) while maintaining competitive accuracy
    \item \textbf{Scaling Advantage}: The power-law efficiency metric demonstrates that EWKAN's architectural innovations enable superior parameter utilization compared to traditional CNN and GNN approaches
\end{itemize}

These results demonstrate that the Kolmogorov-Arnold Network architecture enables unprecedented parameter efficiency in materials property prediction tasks, making it particularly suitable for resource-constrained applications and edge computing scenarios. The NetScore-based evaluation framework provides a robust theoretical foundation for quantifying these efficiency gains.

\clearpage
\section*{Chemical Formula Degeneracy}
While the EWKAN model achieves competitive accuracy in predicting formation energy and electronic properties directly from elemental composition, its architecture—by design—does not incorporate explicit structural descriptors. This choice emphasizes simplicity and generalizability, but can lead to challenges in distinguishing polymorphs that share identical stoichiometry yet differ substantially in atomic configuration and bonding.

\begin{figure}[h]
    \centering
    \includegraphics[width=1\linewidth]{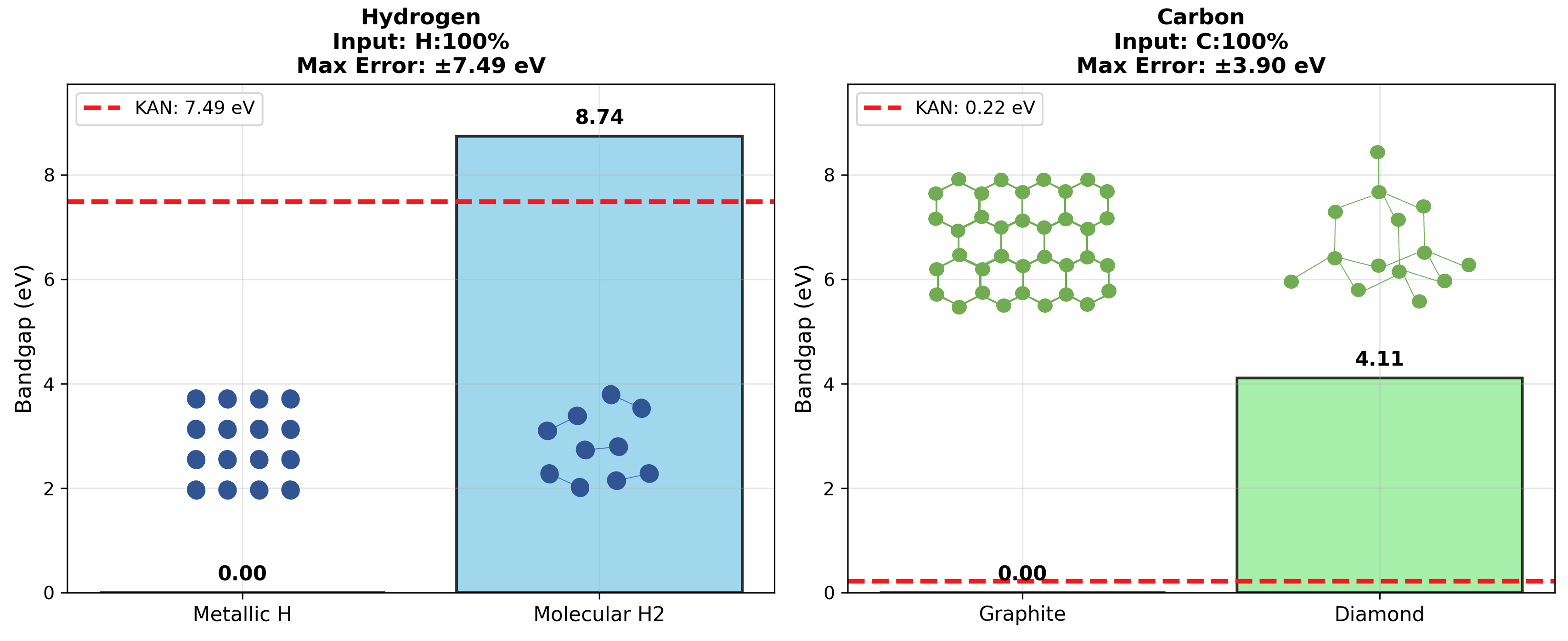}
    \caption{Examples of chemical formula degeneracy in bandgap prediction. Bandgap differences between polymorphs of hydrogen and carbon highlight the EW-KAN model's current limitation in distinguishing structures with identical compositions. Red dashed lines indicate the model's composition-based predictions}
    \label{fig:case_study}
\end{figure}

Chemical formula degeneracy cases, such as graphite versus diamond (pure carbon), and metallic versus molecular hydrogen, highlight this fundamental limitation (see Fig.~\ref{fig:case_study}). Despite having the same chemical formula, these phases differ significantly in bandgap, as computed from first-principles. The EWKAN model, relying solely on elemental composition, yields a single prediction for each input formula, and is thus unable to resolve such structure-induced variations.

\begin{figure}[h]
    \centering
    \includegraphics[width=1\linewidth]{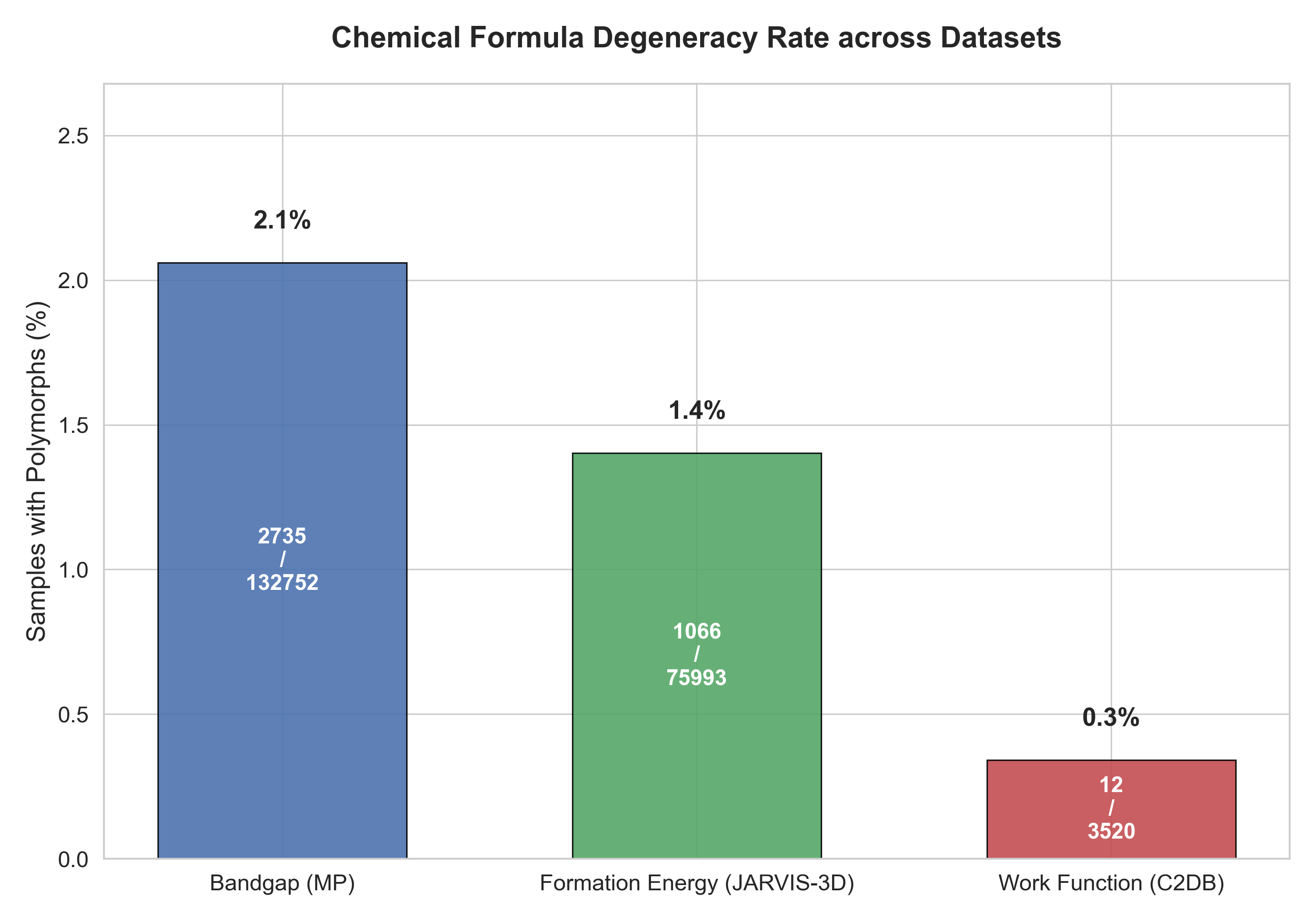}
    \caption{Percentage of samples associated with polymorphic compositions in the three benchmark datasets. Numbers inside the bars indicate the number of polymorphic entries over the total dataset size. Although the fraction is relatively small (2.1\% for band gap, 1.4\% for formation energy, and 0.3\% for work function), these cases highlight scenarios where composition-only models cannot distinguish structural variations.}
    \label{fig:degeneracy_rate}
\end{figure}

To quantify the prevalence of this issue in our datasets, we examine the fraction of entries corresponding to polymorphic compositions. As shown in Fig.~\ref{fig:degeneracy_rate}, the percentage of samples with multiple structural phases is 2.1\% in the band gap dataset (Materials Project subset), 1.4\% in the formation energy dataset (JARVIS-3D), and 0.3\% in the work function dataset (C2DB subset).

\begin{figure}[h]
    \centering
    \includegraphics[width=1\linewidth]{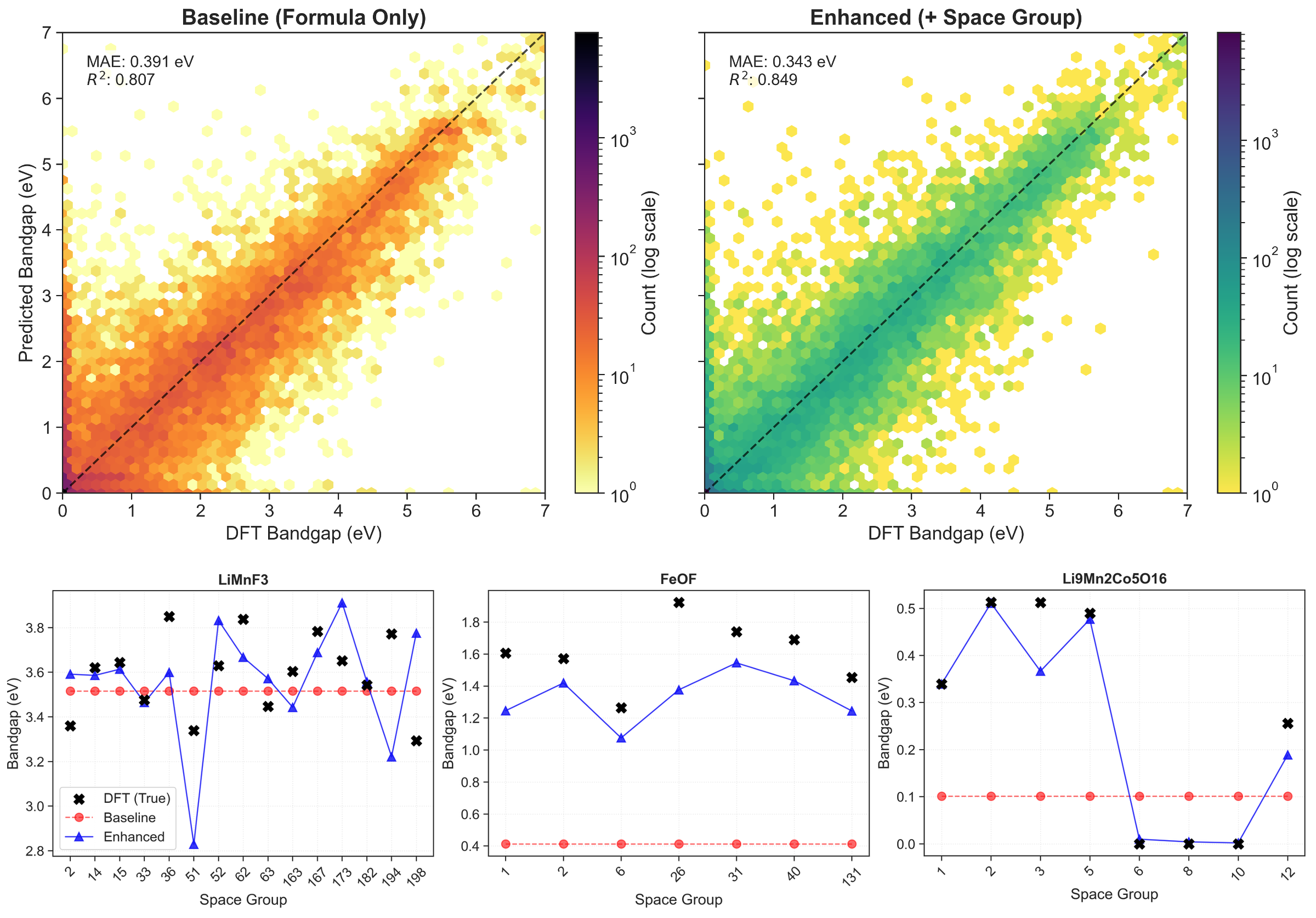}
    \caption{Effect of incorporating space group information on band gap prediction. Top panels: Predicted versus DFT band gaps for the baseline (formula-only) and enhanced (+ space group) models. Bottom panels: Representative polymorphic cases showing band gap variations across different space groups. The baseline model assigns identical predictions to all phases of the same composition, whereas the enhanced model differentiates structures and improves overall accuracy (lower MAE and higher $R^2$).}
    \label{fig:bandgap_sg}
\end{figure}

\begin{figure}[h]
    \centering
    \includegraphics[width=1\linewidth]{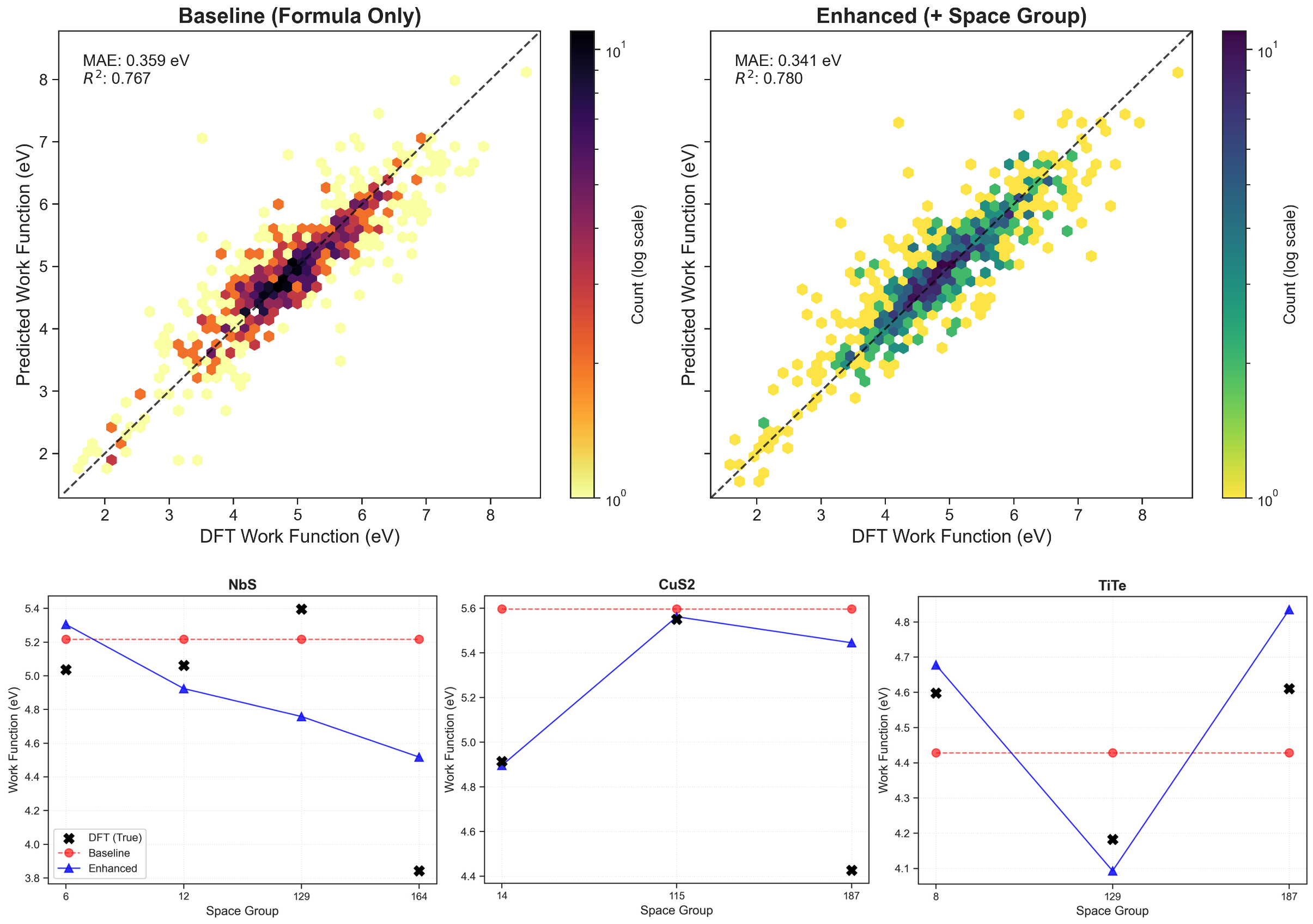}
    \caption{Effect of incorporating space group information on work function prediction. Top panels: Predicted versus DFT work functions for the baseline and enhanced models. Bottom panels: Representative polymorphic materials illustrating structural splitting. Including space group information enables the model to partially resolve composition-induced degeneracy and improves predictive performance.}
    \label{fig:workfunction_sg}
\end{figure}

\begin{figure}[h]
    \centering
    \includegraphics[width=1\linewidth]{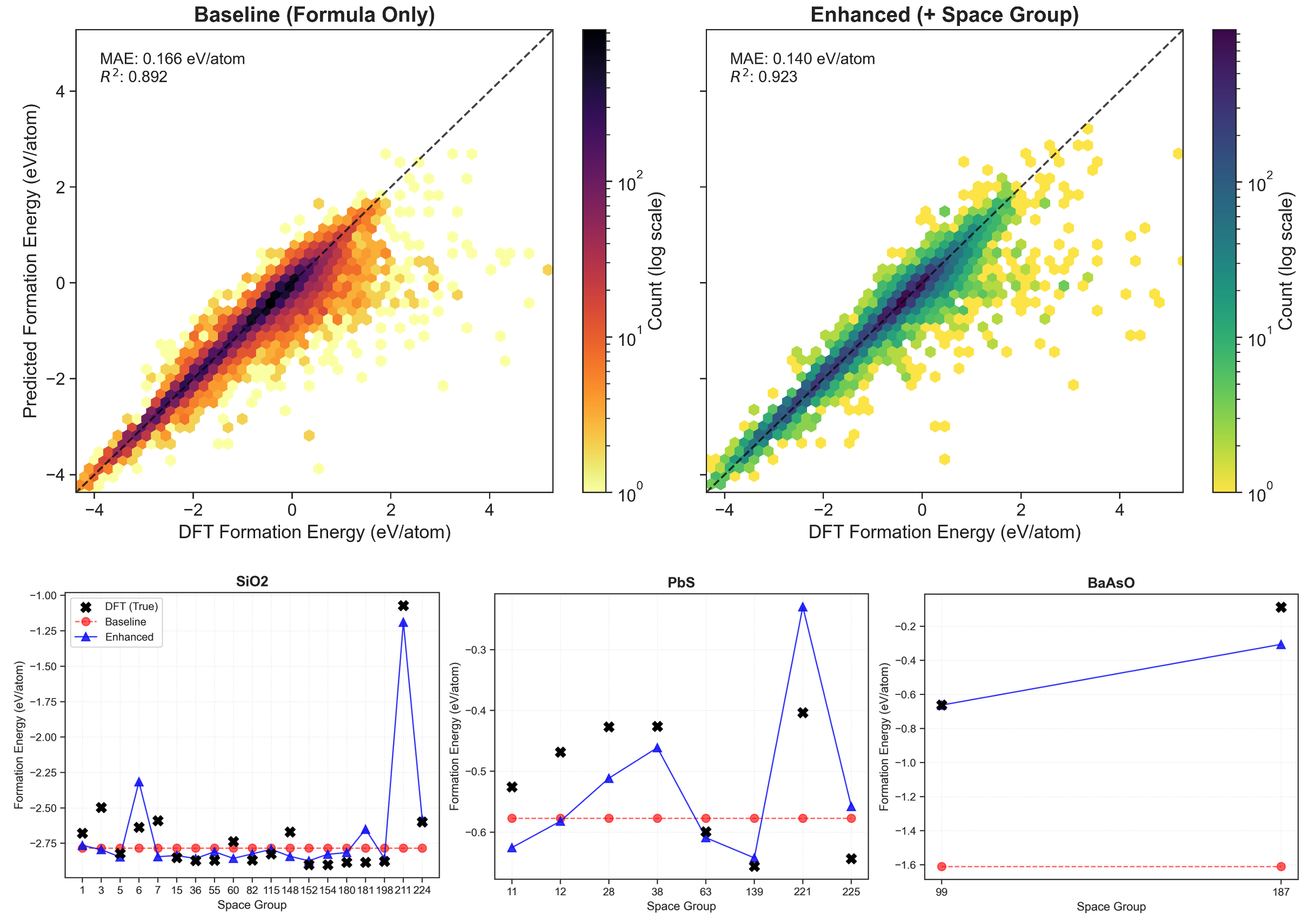}
    \caption{Effect of incorporating space group information on formation energy prediction. Top panels: Predicted versus DFT formation energies for the baseline and enhanced models. Bottom panels: Selected polymorphic compounds demonstrating that the baseline model collapses all phases to a single value, while the enhanced model differentiates space groups and better captures structure-dependent variations.}
    \label{fig:formation_sg}
\end{figure}

To further investigate whether this limitation arises from the KAN architecture or from the composition-only input representation, we incorporate space group information as an additional structural descriptor and retrain the model. Figures~\ref{fig:bandgap_sg}, \ref{fig:workfunction_sg}, and \ref{fig:formation_sg} compare the baseline (formula-only) and enhanced (+ space group) models for band gap, work function, and formation energy, respectively.

In all three properties, adding space group information improves the overall predictive performance, as reflected by reduced MAE and increased $R^2$. More importantly, in representative polymorphic cases (lower panels of each figure), the baseline model assigns identical predictions to all phases of a given composition, whereas the enhanced model differentiates between distinct space groups and captures the corresponding variation in band gap, work function, and formation energy. Even though space group is a coarse structural descriptor, it partially lifts the degeneracy and improves agreement with DFT values.

It is worth noting, however, that this limitation is not intrinsic to the KAN architecture itself, but rather a result of the input representation. The underlying flexibility of KANs provides ample opportunity for enhancement via the inclusion of structural features—such as space group labels, bonding topology, or coordination environment descriptors—into the element-wise processing pipeline. Such integration could allow the model to disambiguate polymorphic materials and better reflect the full diversity of the crystal energy landscape. This suggests a promising future direction for extending KAN-based models toward structure-aware, composition-structure-property prediction frameworks.

\clearpage

\section*{Individual Prediction Check}
\begin{table}[h]
\centering
\caption{Predicted and database formation energies of representative Fe-based compounds.}
\label{tab:fe_compounds}
\begin{tabular}{@{}lll@{}}
\toprule
\textbf{Formula} & \textbf{Predicted (eV)} & \textbf{Database Value (eV)} \\
\midrule
Fe$_3$Sn$_2$ & 0.0270  & 0.0288 \\
Fe$_3$Sn    & 0.0749  & 0.0706, 0.0706, 0.1336, 0.0586, 0.0586 \\
FeO         & -1.0145 & -1.0038, -1.0979, -1.0041, -1.0141, -1.0039, -0.9914, -1.0036, -1.1110, -1.0139 \\
Fe$_2$O$_3$ & -1.3497 & -1.4099, -0.8025, -1.4496, -1.3610, -1.3297, -1.2657 \\
Fe$_3$O$_4$ & -1.2822 & -1.3434, -1.3431, -1.2450, -1.2103, -1.2207, -1.3553, -1.2232 \\
FeS         & -0.3833 & -0.3806, -0.3819, -0.3736, -0.3502, -0.3809, -0.3499, -0.3510, \\
                  &   &     -0.4406, -0.5813, -0.3510, -0.3830, -0.3827, -0.3830 \\
FeS$_2$     & -0.3949 & -0.6353, -0.2779, -0.3816, -0.6273, -0.3816, -0.1779 \\
\bottomrule
\end{tabular}
\end{table}

\end{document}